# The Adoption Paradox for Veterinary Professionals in China: High Use of Artificial Intelligence Despite Low Familiarity

**Authors**: Shumin Li[1], Xiaoyun Lai[2]

**Affiliations:**

[1] College of Veterinary Medicine, Jilin University, Jilin, China 130062

[2] West East Small Animal Veterinary Conference (WESAC), Jiangsu, China 214002

**Corresponding author** :

Shumin Li, DVM, MSc
College of Veterinary Medicine, Jilin University
5333 Xi'an Road, Lvyuan District, Changchun, Jilin, China 130062
shuminli@jlu.edu.cn or aishanglishumin@gmail.com (preferred)
+8618100481863

## Abstract

Introduction: The global integration of artificial intelligence (AI) into veterinary medicine is advancing, yet its adoption in major markets like China remains uncharacterized. This study aimed to provide the first exploratory analysis of AI perception and adoption among veterinary professionals in China.

Methods: A cross-sectional survey was administered to 455 veterinary professionals in China from May to July 2025. Data on AI familiarity, adoption rates, application priorities, and perceived drivers and barriers were analyzed using descriptive statistics and thematic analysis.

Results: We identified a distinct adoption paradox: 71.0% of respondents incorporated AI into their workflow, yet 44.6% of these active users reported low familiarity with the technology. Adoption was primarily practitioner-driven and focused on core clinical tasks, including AI-assisted disease diagnosis (50.1%) and prescription calculation (44.8%). The primary barrier to use was concern about AI reliability and accuracy (54.3%). A strong consensus (93.8%) emerged supporting regulatory oversight of AI by veterinary authorities.

Discussion: The adoption paradox is driven by a practitioner-led, "inside-out" integration model where AI is used to augment clinical capabilities, countered by an "interpretability gap" that limits trust and familiarity. This contrasts with the more administrative, "outside-in" pattern seen in North America. The findings underscore a need for specialized veterinary AI tools, enhanced training focused on critical appraisal, and robust regulatory frameworks to safely harness AI's potential in one of the world's largest veterinary markets.

## Introduction

Over the past decade, there has been extensive research in healthcare demonstrating the potential of artificial intelligence (AI) in healthcare through its capacity to analyze vast datasets, identify complex patterns, and automate routine tasks [1, 2]. Beyond these programmed functions, AI can also yield unintended yet valuable insights, a phenomenon sometimes termed 'opportunistic AI'. For instance, analysis of medical images can reveal secondary pathologies beyond the initial scope of the examination [3, 4]. In veterinary medicine, the promise of AI has been demonstrated through numerous commercial applications that enhance diagnostics, predictive analytics, patient communication, and personalized medicine [5-12]. While the potential of these technologies is widely acknowledged, their practical adoption and impact on clinical decision-making are not yet fully understood, particularly outside of North America and Europe. Evaluating this impact is critical for translating these promising technologies from development into routine clinical practice for both veterinary and human medicine [13].

Recent research, such as the 2024 survey of veterinary professionals in the United States and Canada[14], has provided an initial benchmark, correlating AI familiarity with professional optimism and identifying key implementation barriers in a western context. However, this perspective is geographically limited and leaves a knowledge gap concerning AI adoption in other major international markets. China, with its multi-billion-dollar pet care industry and large network of modern veterinary facilities, represents one of the largest and most dynamic such markets, yet professional attitudes toward and adoption of AI remain completely uncharacterized[15]. Understanding the perceptions, priorities, and challenges within this distinct ecosystem is essential for a complete global assessment of AI's trajectory in the veterinary field.

Therefore, the purpose of this study was to address this knowledge gap by providing the first exploratory analysis of AI perception and adoption among veterinary professionals in China. We hypothesized that the unique demographics, market structure, the legislative frameworks, and technological landscape in China would foster a distinct pattern of AI integration. By characterizing this pattern, we aim to provide crucial insights for technology developers, educators, and regulators seeking to responsibly and effectively support the veterinary profession in China.

## Methods

### *Study Design and Participants*

A descriptive, cross-sectional survey design[16] (Kelley et al., 2003) was used to investigate the adoption and perception of AI among veterinary professionals in China. The study population consisted of veterinary personnel, including veterinarians, technicians, and administrative staff. A convenience sampling approach was employed, recruiting participants between May and July 2025 through two primary channels: in-person at a major national veterinary conference (West East Small Animal Conference (WESAC)) and online through professional networks of the Chinese Veterinary Medical Association (CVMA).

Due to the broad distribution methods employed, which included open professional networks and potential snowball sharing, the exact size of the sampling frame and thus a precise response rate cannot be determined. This is a recognized limitation of web-based surveys designed to achieve wide reach across a large, diffuse population. While this prevents a traditional response rate calculation, the final sample of 455 complete responses provides a robust, novel dataset that meets the target sample size for descriptive analysis (yielding a margin of error of approximately ±5% for proportional estimates) and offers the first quantitative insights into AI adoption in this market. Participation was voluntary and anonymous.

***Survey Instrument***

The data collection instrument was a 26-item questionnaire administered in Mandarin via a commercial online survey platform (Wenjuanxing). The questionnaire was adapted from a previously published 2024 North American (NA) survey[14] to facilitate international contextualization. The adaptation process involved two key steps: 1) Translation and Cultural Validation: The original English questionnaire was translated into Mandarin by a bilingual veterinary professional and then back-translated by an independent translator to ensure conceptual equivalence. 2) Contextual Modification: While the core structure and the majority of questions were retained to allow for contextual comparison, modifications were made to reflect the Chinese market. This included adding examples of AI tools prevalent in China (e.g., Deepseek, DouBao) to the options for Q9 and adjusting demographic categories (e.g., institutional types) to align with the Chinese veterinary landscape. The final instrument included 21 categorical (single or multiple-choice) and 5 open-ended questions, structured into four sections: (1) demographic and professional information; (2) familiarity and usage of AI tools; (3) perceptions, attitudes, and trust in AI; and (4) qualitative feedback. The full questionnaire (original and translated versions) is provided in the Supplementary Material.

A limitation of the survey instrument is the potential ambiguity in Question 8 ("Which AI Tools or Fields Have You Known or Used?"), which conflates awareness with actual use. Furthermore, the combination of "diagnosis" and "treatment" into a single option and the broad scope of "Drug doses and prescription" may limit the granular interpretation of these specific application areas. These limitations are acknowledged and will be considered in the interpretation of the results.

***Statistical Analysis***

All quantitative data were analyzed using Python (3.8.8). Response distributions for all categorical variables were first assessed for non-random patterns using chi-square goodness-of-fit tests. To evaluate the relationships between key variable pairs (e.g., AI familiarity and AI adoption rate), chi-square tests of independence were performed, a method appropriate for analyzing associations between two categorical variables. For all statistically significant associations, Cramér's V was calculated to measure the effect size and determine the strength of the relationship. A p-value of $< 0.001$ was set as the statistical significance threshold for all analyses to control for the increased risk of Type I errors due to multiple comparisons. Qualitative data from open-ended questions were analyzed thematically.

***Contextual Comparison with North American Data***

To situate the findings from China within the global conversation, a descriptive comparison was made with the published summary results of the 2024 NA survey of veterinary professionals (n=3,968) [14]. It is critical to note that this was not a primary aim of the study and that the comparison has inherent constraints. The two surveys were conducted a year apart, used different sampling frames, and most importantly, the professional composition of the cohorts differed fundamentally (e.g., 24.3% veterinarians in NA vs. 81.5% in China). Due to these factors and the lack of access to the raw NA dataset, which prevents adjustment for confounding variables, formal inferential statistical testing between the two cohorts was not performed. Therefore, the comparison is presented as a descriptive overview to highlight potential trends and contextual differences, and any numerical differences should be interpreted with caution.

## Results

### *Demographics and Professional Profile of the Chinese Cohort*

A total of 455 valid responses were collected from veterinary professionals in China. The respondent pool consisted primarily of frontline clinicians, with 81.5% identifying as veterinarians (Figure 1A). Other roles represented included veterinary technicians, assistants, or nurses (5.9%), hospital or clinic managers (4.0%), and veterinary students (2.9%). The cohort was composed of early-to-mid-career professionals. The largest age group was born in the 1990s (57.8%, Figure 1B), followed by those born in the 1980s (24.4%). This aligns with reported professional experience, where 36.5% had 1-5 years of experience and 33.8% had 6-10 years of experience (Figure 1C). The majority of participants (87.2%, Figure 1D) worked in small animal pet hospitals or clinics.

### *Data analysis*

All 21 categorical variables in the Chinese survey demonstrated a significant deviation from a uniform distribution ($p < 0.001$). Chi-square tests of independence on key variable pairs within the cohort revealed that 18 of 45 pairs (40%) had a statistically significant association ($p < 0.001$). The three strongest associations, based on effect size, were between AI Familiarity and AI Adoption Rate (Cramér's $V = 0.412$), AI Adoption Rate and Usage Frequency (Cramér's $V = 0.619$), and perceived Importance of Training and perceived increase in clinic competitiveness from AI (Cramér's $V = 0.415$). (Figure 2). Full visualizations for all response distributions are available in the Supplementary Material.

### *AI Familiarity, Adoption, and Usage in China*

Most Chinese respondents reported low-to-moderate familiarity with AI, with 51.2% describing themselves as "not very familiar" and 4.2% as "completely unfamiliar". Only 9.7% felt "very familiar" and 35.0% felt "somewhat familiar". Despite this, 71.0% of respondents confirmed they have used AI tools in their professional work (Figure 3A). Among AI users (Figure 3B), 19.8% reported using the tools daily and 30.1% used them weekly. Regarding user experience (Figure 3C), 68.8% stated they have "tried some [tools] and are prepared to continue exploring," while 18.9% reported that the tools work well and are used regularly. Usage was dominated by Chinese/locally serviced large language models (LLMs) (Figure 3D), including Deepseek (67.5%) and Doubao (52.5%), while internationally recognized models like ChatGPT and Gemini (20.2%) and veterinary-specific platforms such as MiniVet (20.2%) were also used.

The relationship between user status and self-reported familiarity is detailed in **Table 1**. It shows that adoption rates are highest among those with high familiarity (88.2%), but a substantial number of users (144 out of 323, or 44.6%) report low familiarity ('not very familiar' or 'completely unfamiliar'), directly illustrating the coexistence of high use and low understanding.

**Table 1. Cross-tabulation of AI User Status against Self-Reported Familiarity Levels (n=455).**

| Familiarity Level | Total (n=455) | Users (n=323) | Non-Users (n=132) | Adoption Rate within Group |
|---|---|---|---|---|
| **High Familiarity** | **203** | **179** | **24** | **88.2%** |
| Very Familiar | 44 | 42 | 2 | 95.5% |
| Somewhat Familiar | 159 | 137 | 22 | 86.2% |
| **Low Familiarity** | **252** | **144 (44.6%)** | **108** | **57.1%** |
| Not Very Familiar | 233 | 138 | 95 | 59.2% |
| Completely Unfamiliar | 19 | 6 | 13 | 31.6% |
| **Total** | 455 | 323 (100%) | 132 (100%) | 71.0% |

### *Primary Applications and Perceptions in China*

Respondents primarily used AI for clinical decision-making tasks. The most cited application was AI-assisted disease treatment and diagnosis (50.1%, Figure 4A), followed by calculating and using prescription drugs (44.8%) and treatment planning (36.3%). More administrative functions, such as voice-to-text transcription (25.9%) and automation of electronic health records (20.7%), were cited less frequently. The most perceived benefit of AI was improving efficiency and saving time (86.6%, Figure 4B), followed by improving the accuracy of diagnosis (56.7%), reducing administrative workload (49.2%) and medical record related work (39.6%). While 45.1% of respondents (Figure 4C) expressed confidence in AI, 30.3% (Figure 4D) stated they do not trust unverified AI output for diagnosis.

### *Barriers and Drivers for Future Adoption in China*

The primary barrier to adoption was concern about AI reliability and accuracy (54.3%, Figure 5A). The second-most cited barrier was a lack of training and knowledge (49.5%). Consequently, the top-ranked factor that would promote usage was more training opportunities (59.1%, Figure 5B). A consensus of 93.8% of respondents (Figure 5C) believe AI use should be regulated by veterinary authorities, with opinion split between needing "strict regulation" (46.6%) and regulation that "should retain some flexibility" (47.2%).

### *Descriptive Comparison with North American Data*

When viewed alongside the previously published NA data, notable differences in professional composition and adoption patterns are observed (**Table 2**). The Chinese cohort consisted of a higher proportion of veterinarians (81.5% vs 24.3%) and reported a higher rate of AI adoption (71.0% vs 39.2%). Application priorities also differed. The Chinese cohort more frequently prioritized clinical tasks like diagnosis and disease detection (50.1% vs. 34.1%) and prescription calculation (44.8% vs. 17.6%). In contrast, the NA cohort more frequently reported using AI for administrative tasks like record-keeping (39.0% vs. 20.7%) and imaging and radiology (39.0% vs. 24.8%). Concerns about reliability and accuracy were the top barrier in both groups (NA: 70.3%, China: 54.3%). NA professionals rated data security and privacy (53.9% vs. 26.6% in China), cost (42.6% vs. 20.4% in China), and fear of job displacement (36.1% vs. 20.9% in China) as more prominent barriers than their Chinese counterparts.

**Table 2**. Descriptive Comparison of Key Metrics between North American (2024) and Chinese (2025) Veterinary Professionals. Note, direct statistical comparison is not appropriate due to significant differences in cohort composition and methodology. Data should be interpreted as contextual trends.

| Comparison | North America (n=3,968) | China (n=455) |
|---|---|---|
| **Composition** | | |
| Veterinarian | 24.3% | 81.5% |
| Tech/Assistant | 37.7% | 5.9% |
| Vet Student | 13.6% | 2.9% |
| Manager/Exec | 11.0% | 4.0% |
| Receptionist | 6.7% | 1.3% |
| Other | 6.6% | 4.4% |
| **Adoption & Familiarity** | | |
| AI Adoption Rate | 39.2% | 71.0% |
| AI Faimilarity (Very/Somewhat) | 83.8% | 44.6% |
| **Primary Applications** | | |
| Diagnosis and disease detection | 34.1% | 50.1% |
| Drug dosages and prescription | 17.6% | 44.8% |
| Treatment Planning | 31.1% | 36.3% |
| Imaging and radiology | 39.0% | 24.8% |
| Record-keeping & admin tasks | 39.0% | 20.7% |
| Voice-to-text transcription | 36.9% | 25.9% |
| Client communication & education | 31.7% | 35.6% |
| **Barriers** | | |
| Reliability and accuracy | 70.3% | 54.3% |
| Data security and privacy | 53.9% | 26.6% |
| Cost of implementation | 42.6% | 20.4% |
| Lack of sufficient tool options | Unclear | 36.0% |
| Lack of training and knowledge | 42.9% | 49.5% |
| Regulatory or legal issues | 42.1% | 31.2% |
| Fear of job displacement | 36.1% | 20.9% |

## Discussion

This study provides the first exploratory analysis of AI adoption among veterinary professionals in China, revealing a distinct integration approach. The central finding is an "adoption paradox": while adoption is naturally highest among those with high familiarity (88.2%), our data reveals that 44.6% of active users report low familiarity with the technology. The co-existence of high usage and low familiarity suggests that this market is undergoing rapid, bottom-up technological assimilation. It paints a portrait of a profession in the early, enthusiastic, yet cautious stages of adoption, where practice is rapidly outpacing formal proficiency and trust. This "adoption paradox" helps us understand the unique dynamics of AI integration in China's veterinary sector (companion animal care in particular).

The high adoption rate observed in this study is propelled by a powerful confluence of user demographics and systemic needs. The first driver is the demographic profile of the sampled respondents. Our cohort was predominantly young, with 57.8% born in the 1990s, which may reflect a broader trend of a youthful veterinary workforce in China and positions them as digital natives who are naturally inclined to leverage technology for problem-solving. They proactively experiment with accessible, general-purpose tools like large LLMs (e.g., Deepseek, used by 67.5% of respondents) to streamline their clinical workflow. This represents a "bottom-up" adoption model, driven by individual practitioners rather than institutional mandate. The second, and arguably more profound, driver is a systemic need rooted in the structure of the veterinary profession in China. In the United States and Canada, veterinary medicine is a postgraduate profession requiring a doctoral degree (Doctor of Veterinary Medicine), ensuring a high and consistent standard of foundational clinical expertise. The system is additionally supported by a robust and accessible network of board-certified specialists [17]. Whereas the pathway to veterinary licensure in China typically begins with a bachelor's or even an associate degree from a vocational school[18-20]. This results in a larger, more demographically varied professional base whose members may have little initial clinical training and, crucially, more limited access to board-certified specialists. In this context, AI is not merely a convenient tool for saving time; it evolves into a necessary form of digital decision support. However, this reliance carries inherent risks. In a professional landscape with variable baseline training, there is a heightened danger of uncritical acceptance of AI outputs, often termed "automation bias. " It is critical to emphasize that AI serves as a "digital clinical ally" to augment decision-making, not a replacement for the critical appraisal required for diagnosis and treatment. The ultimate professional responsibility remains with the clinician to verify AI suggestions against clinical evidence.

This drive for clinical augmentation is clearly reflected in the high prioritization of core clinical applications—AI-assisted disease diagnosis (50.1%), prescription calculation (44.8%), and treatment planning (36.3%)—which reveals an "inside-out" strategy. Here, AI is being leveraged from the very core of clinical practice to address potential knowledge gaps, mitigate the unevenness of foundational training, and standardize the quality of care across a diverse pet care landscape. For many practitioners, an AI diagnostic assistant is a powerful ally offering accessible second opinions, yet one that requires vigilant oversight given the potential for hallucination or error.

If AI is so integral to practice, what then explains the pervasive low familiarity and trust reported by these same users? The answer lies in the nature of the tools being used and the resulting "interpretability gap." The current wave of AI adoption in China is dominated by general-purpose LLMs. While highly capable, these models often function as "black boxes" [21,22]; they provide outputs without transparent, step-by-step reasoning that a clinician can easily follow and verify. This lack of explainability is the fundamental barrier to both familiarity and trust. A veterinarian can use an LLM to generate a list of differential diagnoses but cannot be certain of the underlying clinical logic and the information source, potentially missing a crucial, low-probability but high-stakes condition. This erodes confidence and perpetuates a state of low familiarity, in other words, practitioners are actively utilizing the tools in their workflows but do not feel they understand the underlying technology or its reliability boundaries. This interpretability gap is reflected in the data as well. While 45.1% of respondents expressed general confidence in AI, a substantial 30.3% of respondents stated they do not trust unverified AI output for diagnosis. Although our study's demographic homogeneity (predominantly early-career) precludes a granular statistical comparison of trust across experience levels, the nature of this 'interpretability gap' suggests a differential impact. For the experienced, it breeds caution; for the less trained, it risks being overlooked in favor of utility. This sentiment was reinforced by qualitative comments describing tools as "sometimes wrong" or "too basic". Therefore, the high usage exists within a carefully bounded comfort zone. Practitioners are willing to deploy AI for tasks where the output is verifiable (e.g., dose calculation, where a formula can be checked) or for generating initial ideas that will be scrutinized. However, they are hesitant to cede authority to AI for complex, high-stakes clinical decisions. This gap between using AI as an auxiliary tool and relying on it as a facilitated diagnostic partner is the crux of the adoption paradox. The profession, especially the experienced ones, is caught between recognizing AI's immense utility and being aware of its current limitations and risks.

When the patterns observed in China are compared with the established data from North America, it becomes evident that the trajectory of a global technology is shaped by local ecosystems. The comparison reveals two different adoption narratives, each logically consistent with its own context. The North American model, as reported in the 2024 survey, can be characterized as a cautious, "outside-in," and provider-driven approach. With a highly standardized, postgraduate-trained workforce and a robust network of specialists, the most pressing daily challenges are often operational workload and rising administrative costs. Consequently, the initial integration of AI has logically focused on administrative tasks like record-keeping (39.0%) and voice-to-text transcription (36.9%). This is a strategy that initially targets areas with less direct clinical liability, building a foundation of efficiency upon which more advanced clinical tools might later be integrated.

In contrast, the Chinese model is a need-driven, "inside-out," and practitioner-led phenomenon. The different participant composition, 81.5% veterinarians in China versus 24.3% in NA, is a key explanatory variable. This clinician-heavy sample, operating within an ecosystem of varied training and limited specialist access, has pivoted directly to using AI for clinical augmentation. This divergence demonstrates that adoption priorities are a direct reflection of local workforce demographics, educational structures, market maturity, and underlying professional needs. The trajectory of AI is not a monolithic, one-size-fits-all global rollout but a mosaic of regional adaptations.

The findings of this study carry significant implications for the global veterinary community. The strong consensus (93.8%) among Chinese respondents calling for regulatory oversight, coupled with their anxieties about liability and skill degradation, signals an urgent need for action. Professional associations and regulatory bodies should proactively develop standards for AI tool validation, algorithmic transparency, and clear liability frameworks. The question of "who is responsible? The doctor or the AI?" must be answered to prevent a chilling effect on adoption or, conversely, to prevent over-reliance on unvalidated tools [23]. For educators, the data underscore a universal need to evolve veterinary curricula [24, 25]. Moving beyond basic computer literacy, the next generation of veterinarians must be equipped with skills in data science and the ability to critically appraise AI technologies. The inclusion of courses focused on digital health ethics and AI literacy is a vital step in this direction. For technology developers, they should be aware that a new generation of veterinary-specific AI tools is needed. The demand for more specialized tool options and the interpretability gap create a clear market opportunity for solutions that are explainable, validated on veterinary data, and seamlessly integrated into clinical workflows.

Looking forward, this study suggests that global AI adoption will continue along multiple, parallel tracks. In mature markets, integration will likely remain systematic and caution-led, while in emerging markets, practitioner-driven adoption could accelerate, making it possible to leapfrog certain stages of development but also carrying heightened risks that require vigilant management. Future research must employ longitudinal designs to track this evolution and qualitative methods to thoroughly explore the clinical reasoning and decision-making processes of veterinarians using AI. Furthermore, studies investigating client acceptance and the economic models of AI-assisted services will be critical to understanding its ultimate role and sustainability in veterinary practice.

This study has several limitations that must be considered when interpreting the results. First, the use of a convenience sampling method, despite its strengths in reaching a broad national sample, may introduce selection bias [26]. The recruitment through major conference and professional association channels likely overrepresents professionals who are more engaged with continuing education and potentially more technologically adept, which may inflate the reported adoption rates and familiarity. The inability to calculate a precise response rate due to the broad and diffuse nature of the sampling frame further complicates the assessment of this bias and the generalizability of the findings to the entire veterinary population in China. Second, the potential for ambiguity in key survey questions, particularly Question 8 ("Which AI Tools or Fields Have You Known or Used?"), which conflates awareness with actual use, may lead to an overestimation of adoption for specific applications. Finally, while the descriptive comparison with North American data provides valuable international context, it is constrained by fundamental differences in cohort composition, sampling timeframes, and methodology. The lack of access to the raw NA dataset precluded any multivariable analysis to adjust for confounding factors, meaning the observed differences should be interpreted as generating hypotheses about regional variation rather than confirming them.

Furthermore, while the demographic profile of our respondents (e.g., a high proportion of young, small-animal veterinarians) is consistent with reported trends in China's urban veterinary sector[27,28], the convenience sampling method may not fully capture the diversity of the national veterinary workforce, including those in rural areas or large animal practice. This limits the generalizability of our findings to the entire veterinary population in China.

In conclusion, despite its limitations, this study provides a snapshot of AI adoption in a global market. The integration of AI into veterinary medicine is a complex, socially embedded process. The adoption paradox in China, high use amid low familiarity, is an example of this complexity. By illuminating this unique pathway, our study moves the conversation beyond simple adoption metrics and into the richer territory of how and why technology is adopted in different contexts. For all stakeholders, from clinicians and educators to developers and regulators, success will depend on recognizing and responding to these distinct regional narratives. By doing so, the global veterinary community can responsibly guide the development of AI, ensuring it truly advances animal health and empowers, rather than replaces, the veterinary professionals dedicated to it.

Figure legends:

**Figure 1.** Demographic characteristics of study participants. Bar charts show the distribution of responses to four survey questions: (A) Your Role in Veterinary Field, (B) Age, (C) Your Work Experience in Veterinary Field, and (D) Type of Institution You Work At. Percentages for each category are displayed on the bars, along with the number of respondents (n) for that category.

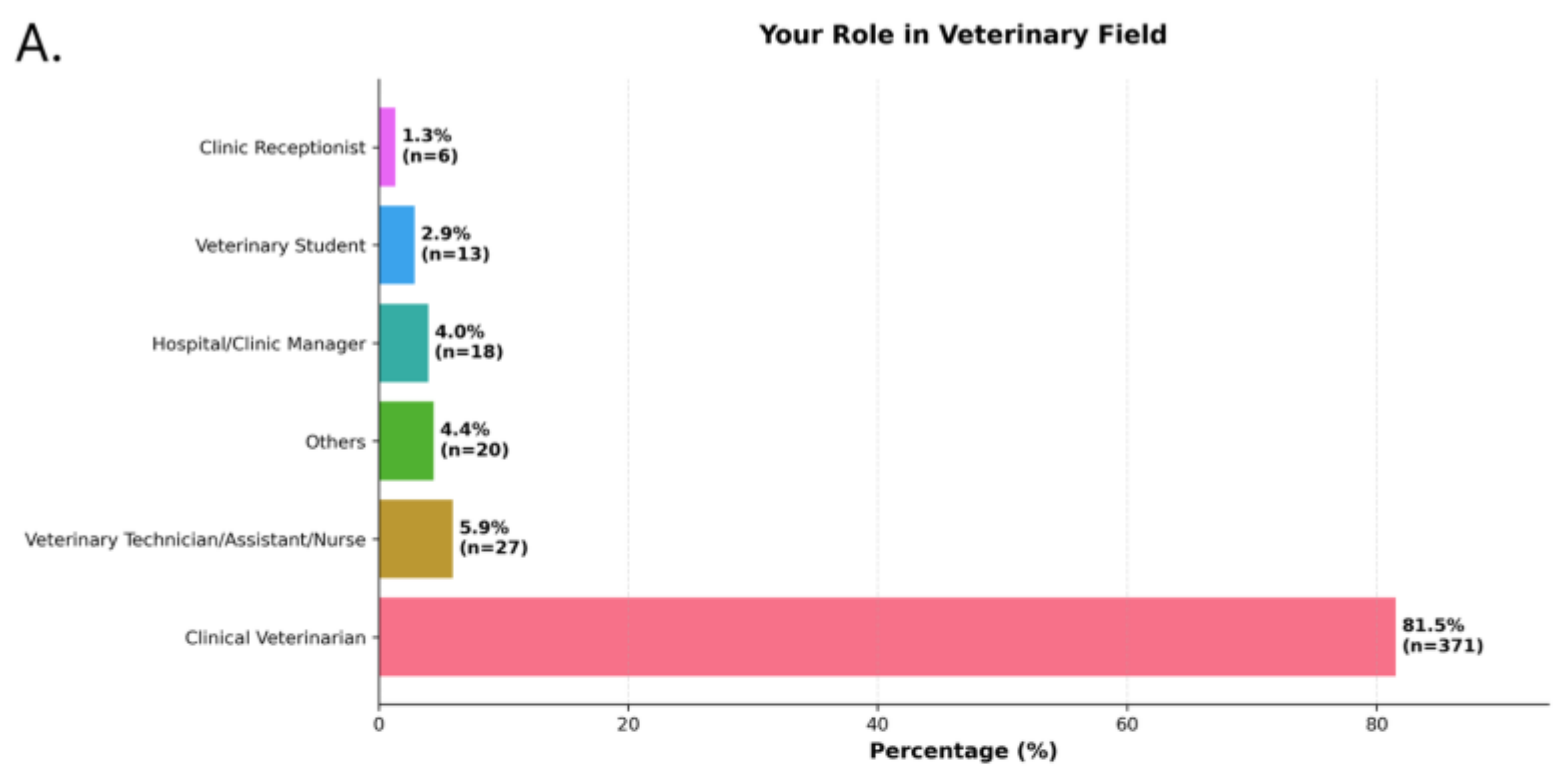
A.
Your Role in Veterinary Field
Clinic Receptionist
Veterinary Student
Hospital/Clinic Manager
Others
Veterinary Technician/Assistant/Nurse
Clinical Veterinarian
1.3% (n=6)
2.9% (n=13)
4.0% (n=18)
4.4% (n=20)
5.9% (n=27)
81.5% (n=371)
0
20
40
60
80
Percentage (%)

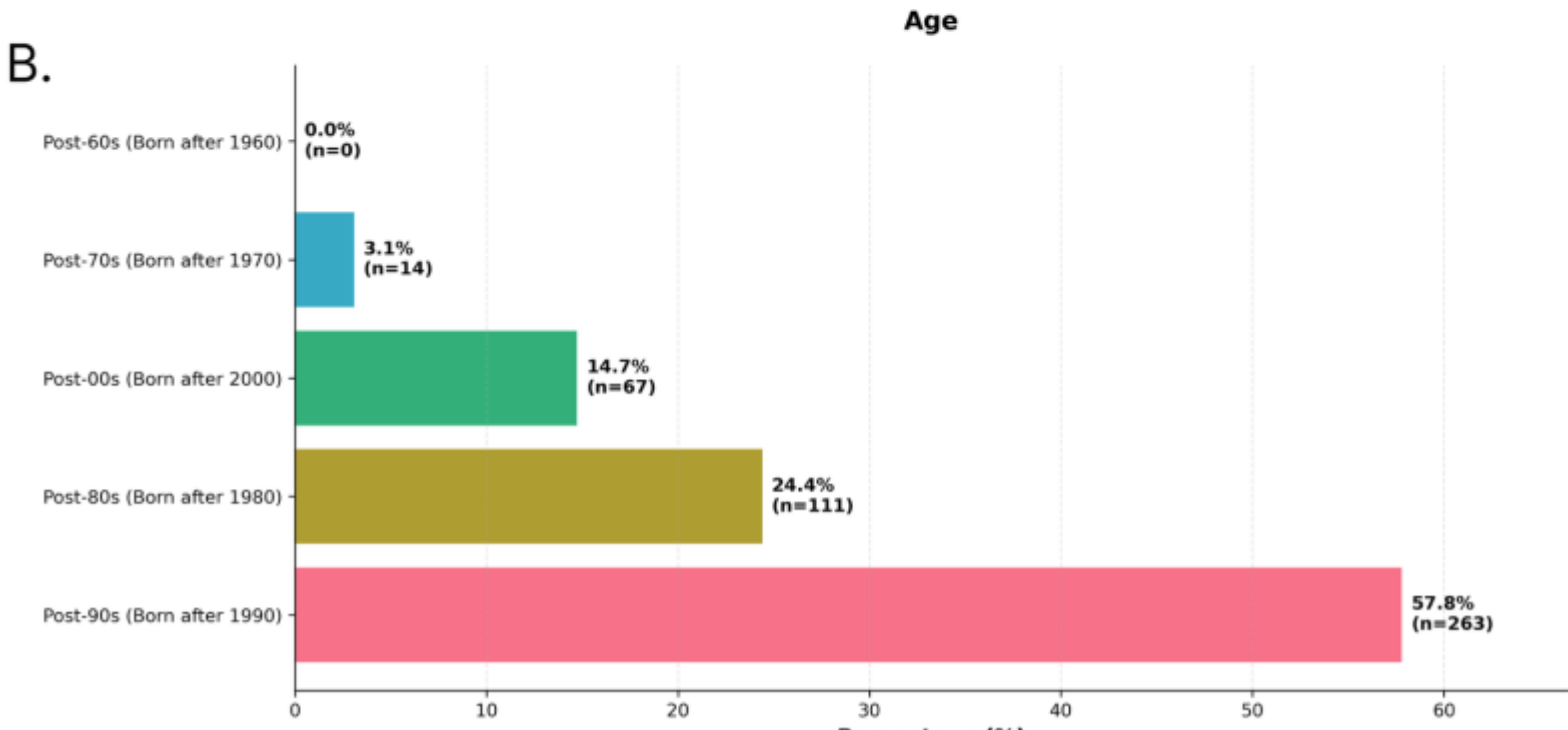
B.
Age
Post-60s (Born after 1960)
Post-70s (Born after 1970)
Post-00s (Born after 2000)
Post-80s (Born after 1980)
Post-90s (Born after 1990)
0.0% (n=0)
3.1% (n=14)
14.7% (n=67)
24.4% (n=111)
57.8% (n=263)
0
10
20
30
40
50
60
Percentage (%)

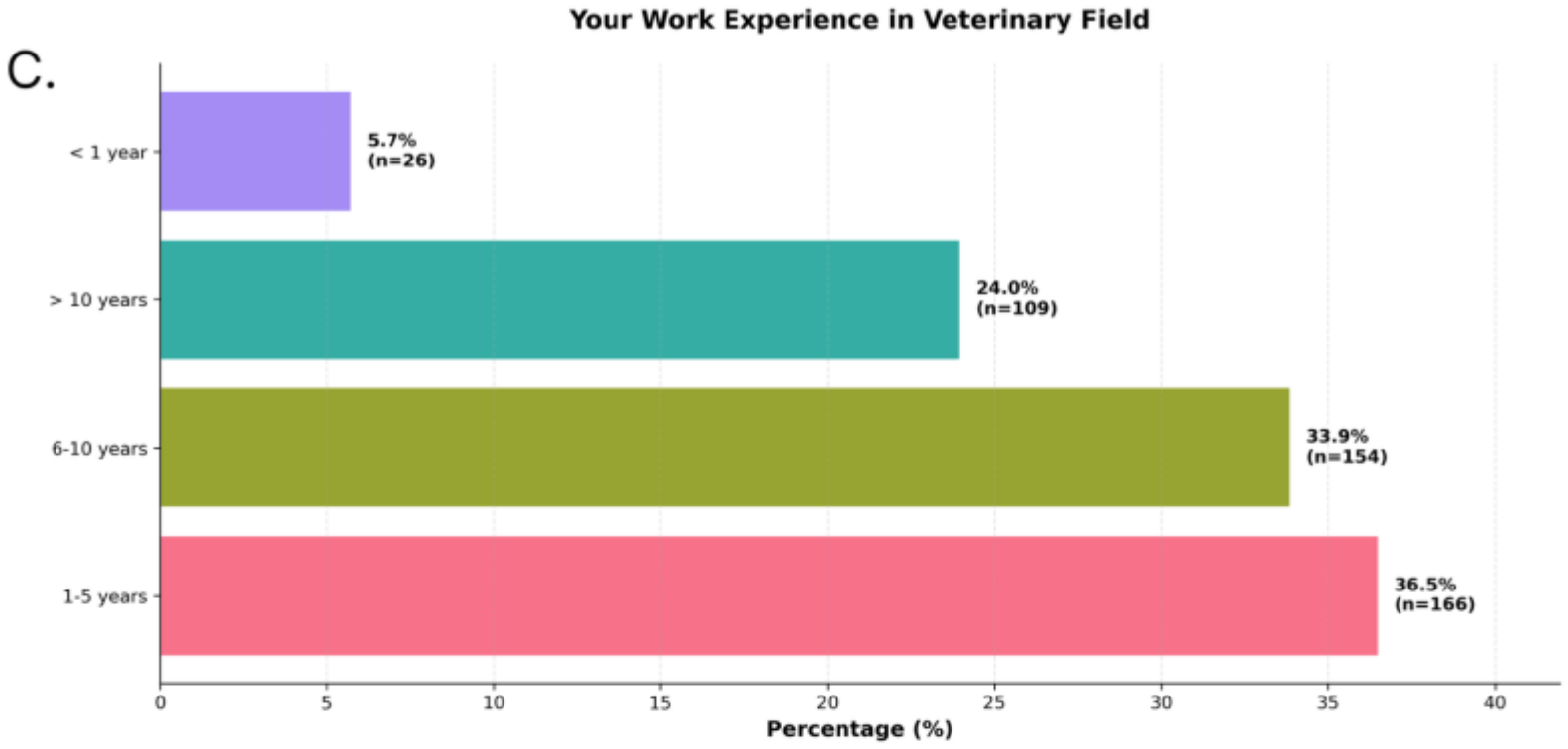
C.
Your Work Experience in Veterinary Field
< 1 year
> 10 years
6-10 years
1-5 years
5.7% (n=26)
24.0% (n=109)
33.9% (n=154)
36.5% (n=166)
0
5
10
15
20
25
30
35
40
Percentage (%)

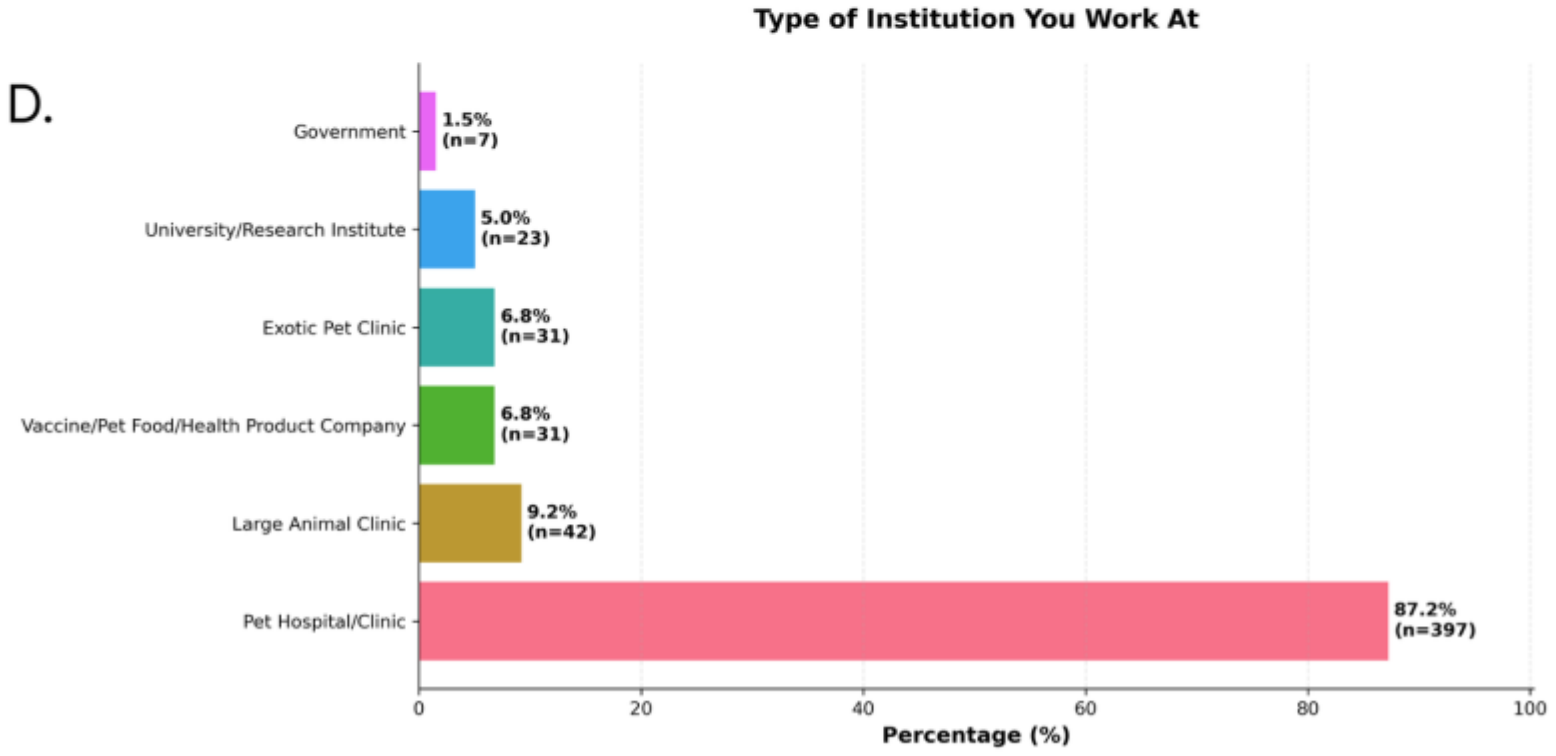
D.
Type of Institution You Work At
Government
University/Research Institute
Exotic Pet Clinic
Vaccine/Pet Food/Health Product Company
Large Animal Clinic
Pet Hospital/Clinic
1.5% (n=7)
5.0% (n=23)
6.8% (n=31)
6.8% (n=31)
9.2% (n=42)
87.2% (n=397)
0
20
40
60
80
100
Percentage (%)

**Figure 2**. Heatmap illustrating the association strength between key categorical variables using Cramér's V from Chi-square tests of independence. The intensity of the blue color and the numerical value in each cell represent the effect size of the association between the corresponding pair of variables. Of the 45 variable pairs tested, 18 (40%) demonstrated a statistically significant association ($p<0.001$). The most notable strong associations are between AI Adoption Rate and Usage Frequency ($V=0.619$), Training Importance and Clinic Competitiveness ($V=0.415$), and AI Familiarity and AI Adoption Rate ($V=0.412$).

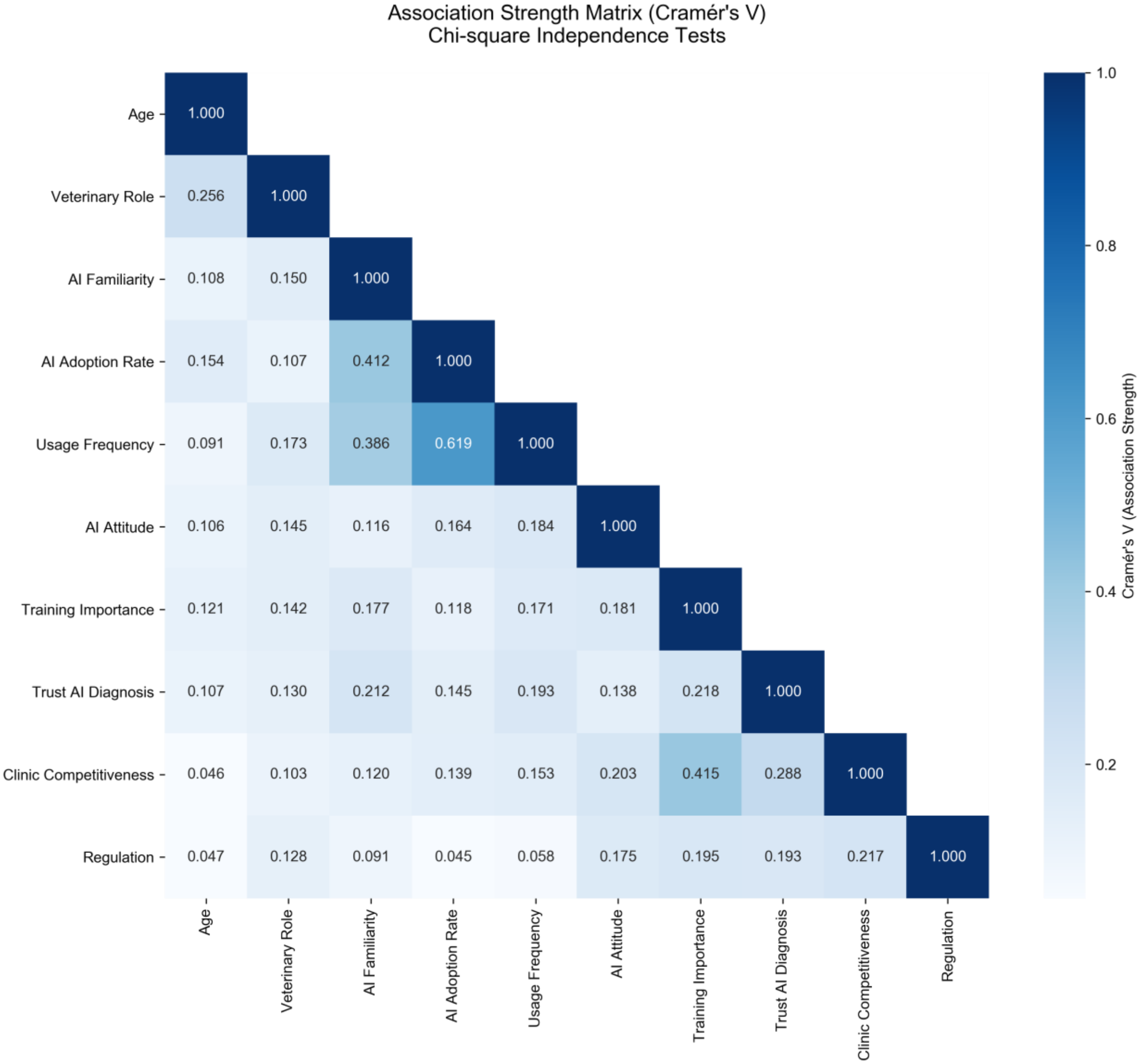

**Figure 3**. Survey results that highlight AI familiarity, adoption, and usage patterns in China. Bar charts show the distribution of responses to five survey questions: (A) Have You Used AI Tools in Veterinary Practice, (B) Frequency of AI Tool Usage at Work, (C) Your Experience with AI Tools,, and (D) What AI Tools Have You Used? Percentages for each category are displayed on the bars, along with the number of respondents (n) for that category.

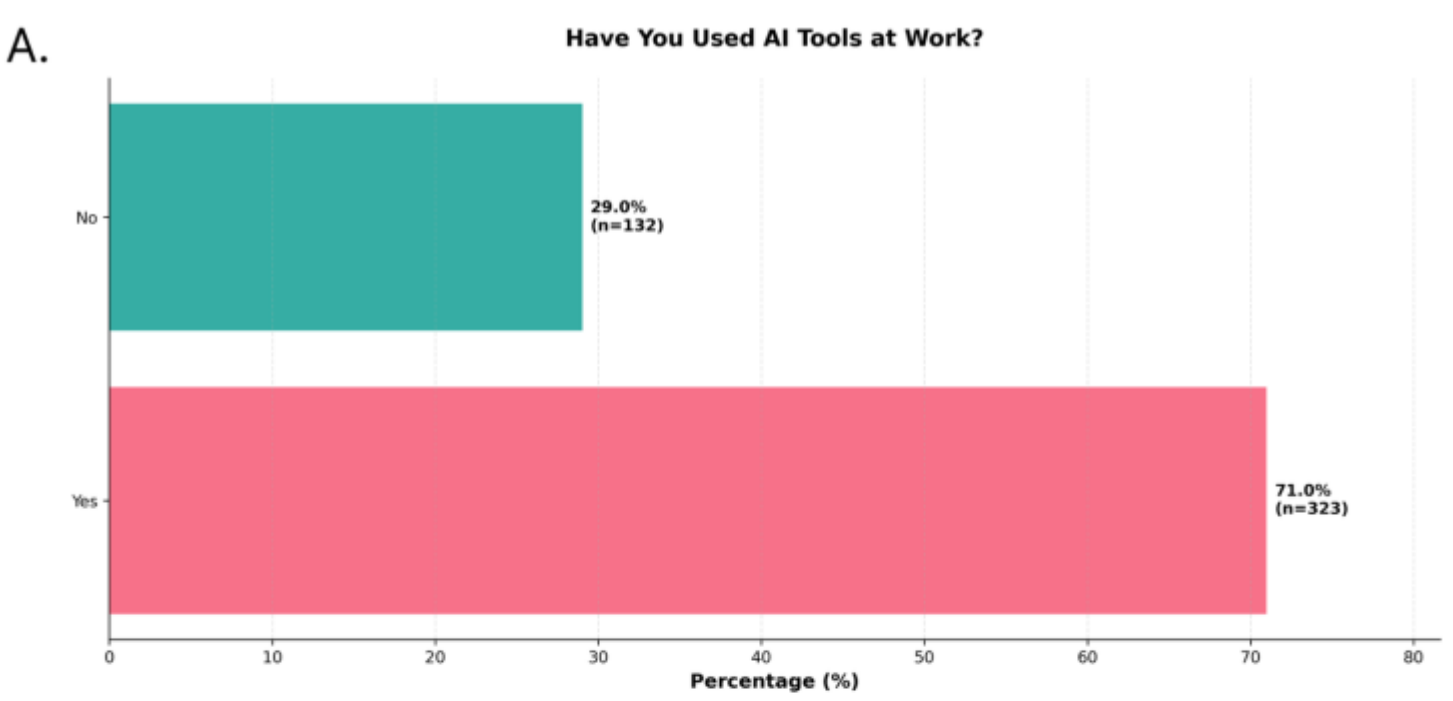
A.
Have You Used AI Tools at Work?
No
29.0%
(n=132)
Yes
71.0%
(n=323)
0
10
20
30
40
50
60
70
80
Percentage (%)

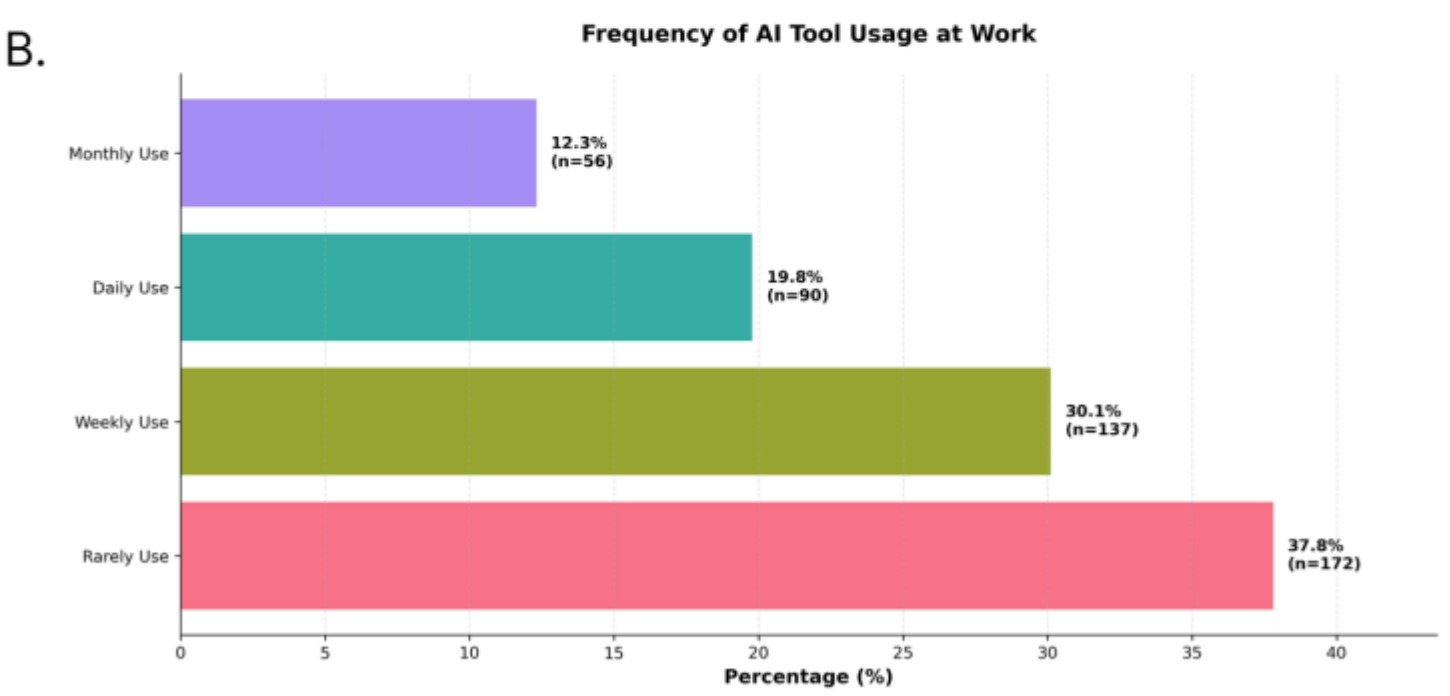
B.
Frequency of AI Tool Usage at Work
Monthly Use
12.3%
(n=56)
Daily Use
19.8%
(n=90)
Weekly Use
30.1%
(n=137)
Rarely Use
37.8%
(n=172)
0
5
10
15
20
25
30
35
40
Percentage (%)

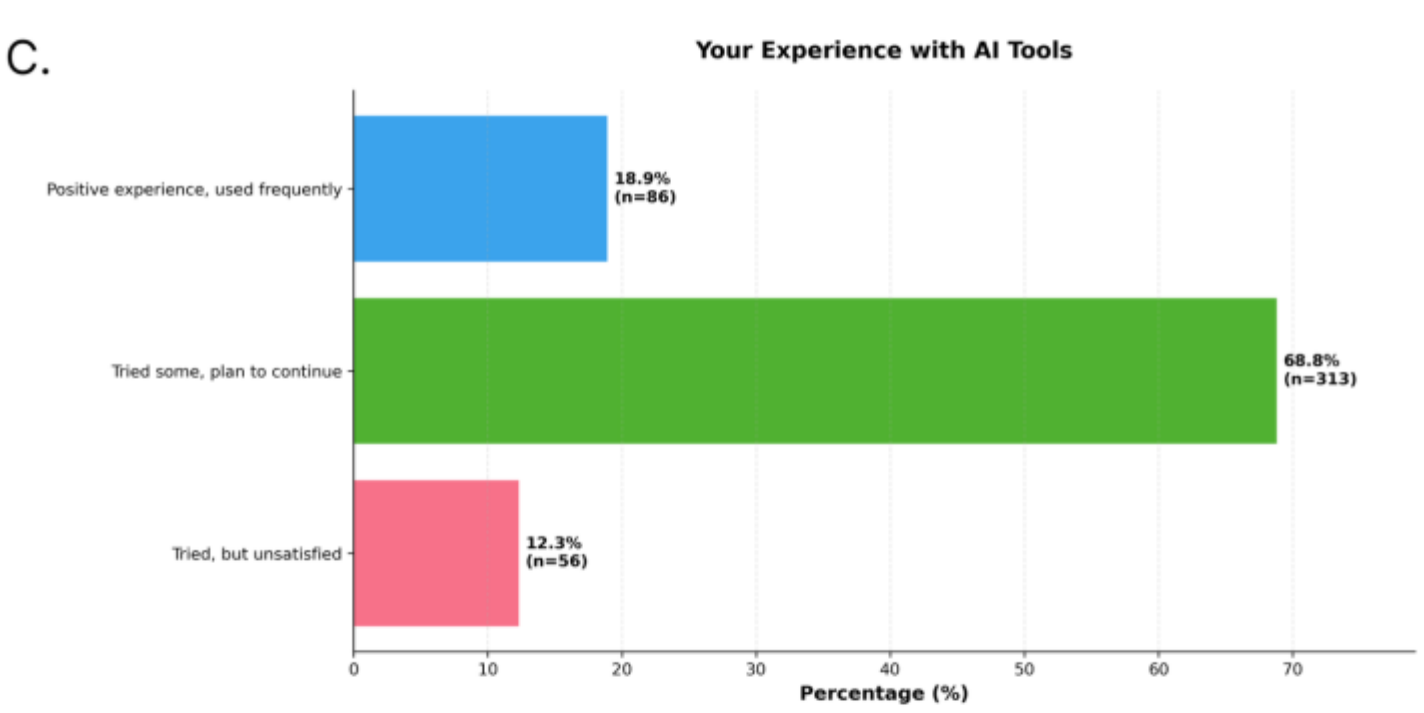
C.
Your Experience with AI Tools
Positive experience, used frequently
18.9%
(n=86)
Tried some, plan to continue
68.8%
(n=313)
Tried, but unsatisfied
12.3%
(n=56)
0
10
20
30
40
50
60
70
Percentage (%)

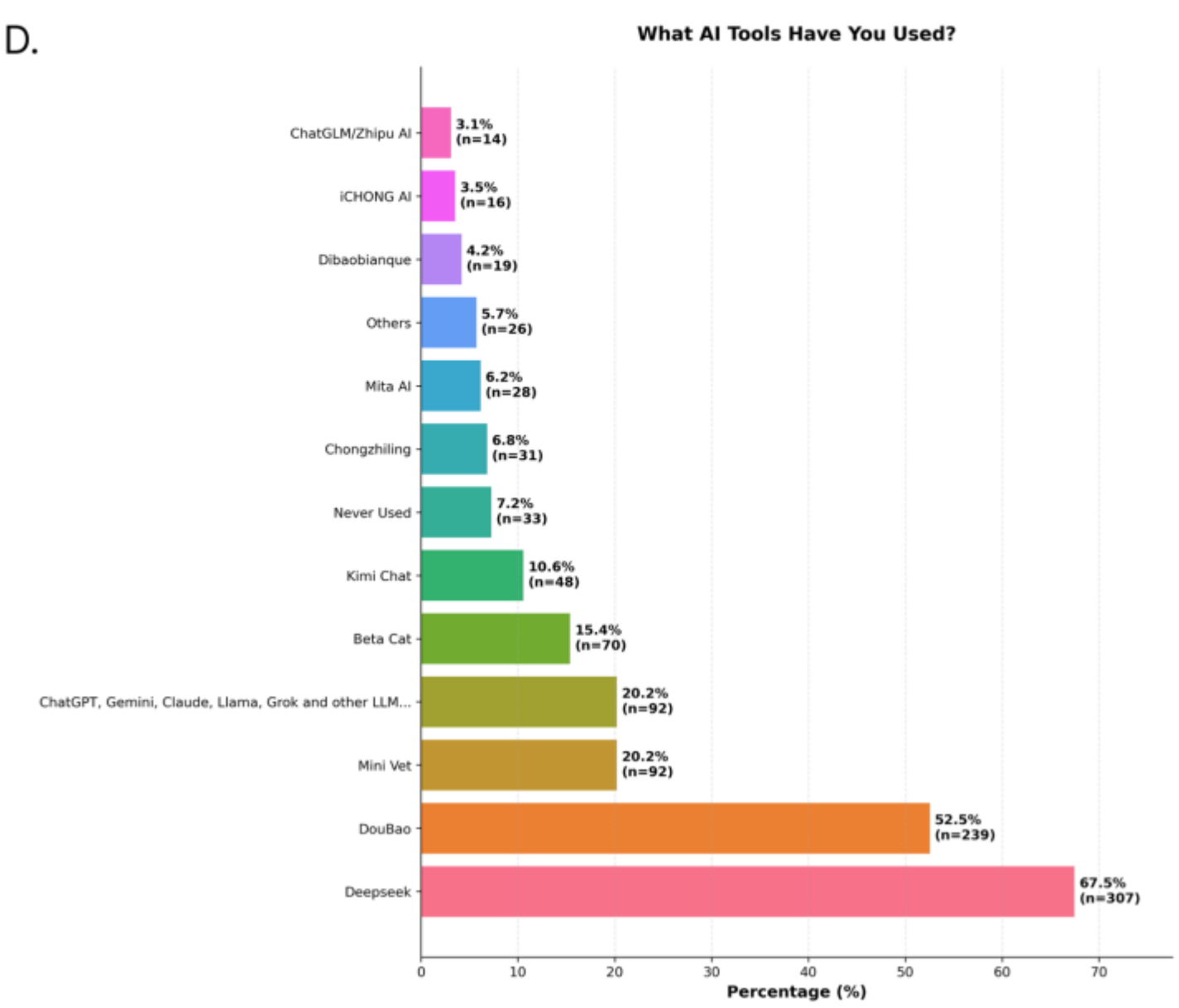
D.
What AI Tools Have You Used?
ChatGLM/Zhipu AI
3.1%
(n=14)
iCHONG AI
3.5%
(n=16)
Dibaobianque
4.2%
(n=19)
Others
5.7%
(n=26)
Mita AI
6.2%
(n=28)
Chongzhiling
6.8%
(n=31)
Never Used
7.2%
(n=33)
Kimi Chat
10.6%
(n=48)
Beta Cat
15.4%
(n=70)
ChatGPT, Gemini, Claude, Llama, Grok and other LLM...
20.2%
(n=92)
Mini Vet
20.2%
(n=92)
DouBao
52.5%
(n=239)
Deepseek
67.5%
(n=307)
0
10
20
30
40
50
60
70
Percentage (%)

**Figure 4**. Survey results that highlight perceived benefits, attitudes, and trust in AI use in China. Bar charts show the distribution of responses to four survey questions: (A) AI Tools or Fields You Know or Have Used?, (B) Main Benefits AI Brings to Veterinary Field, (C) Your Attitude Toward AI Applications in Veterinary Field, and (D) Do You Trust AI Diagnosis Without Human Verification? Percentages for each category are displayed on the bars, along with the number of respondents (n) for that category.

A.

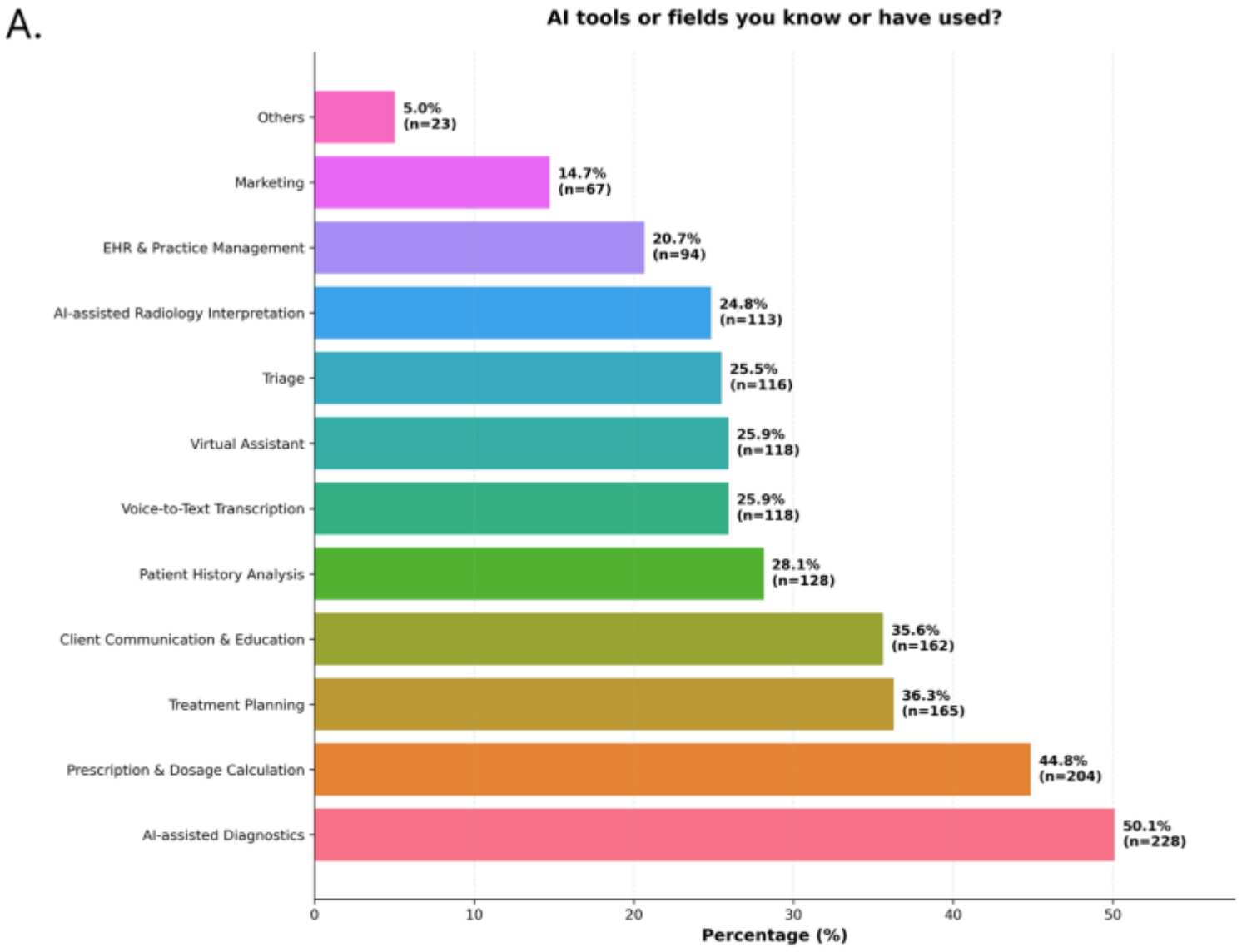

B.

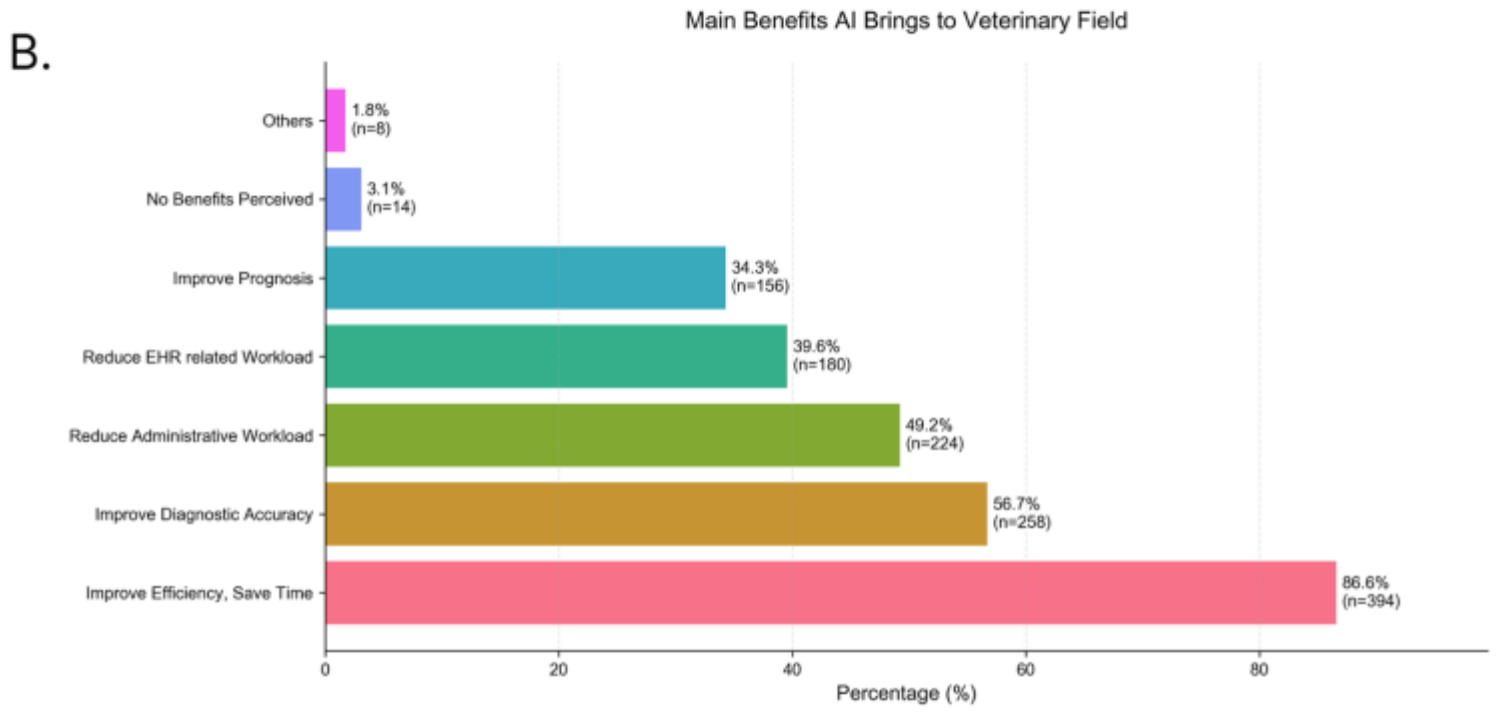

C.

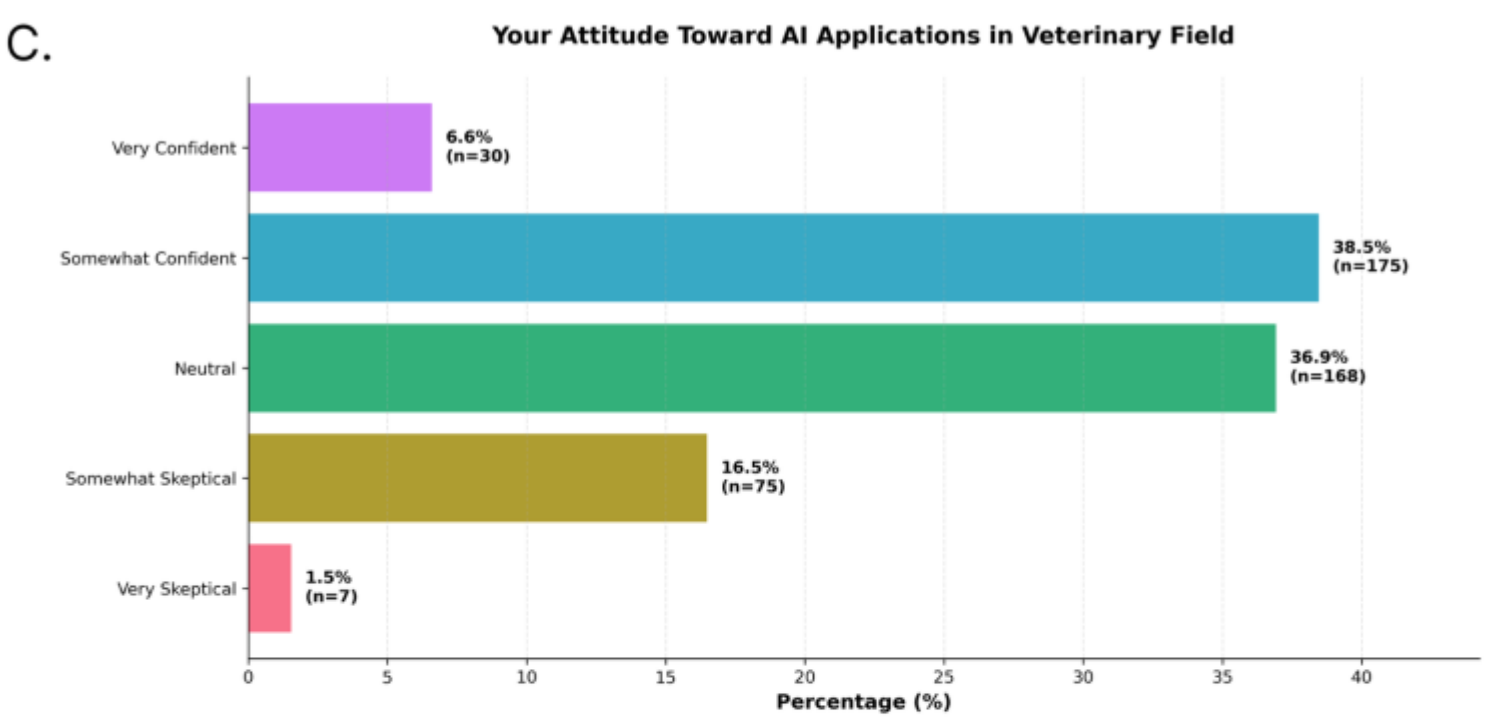

D.

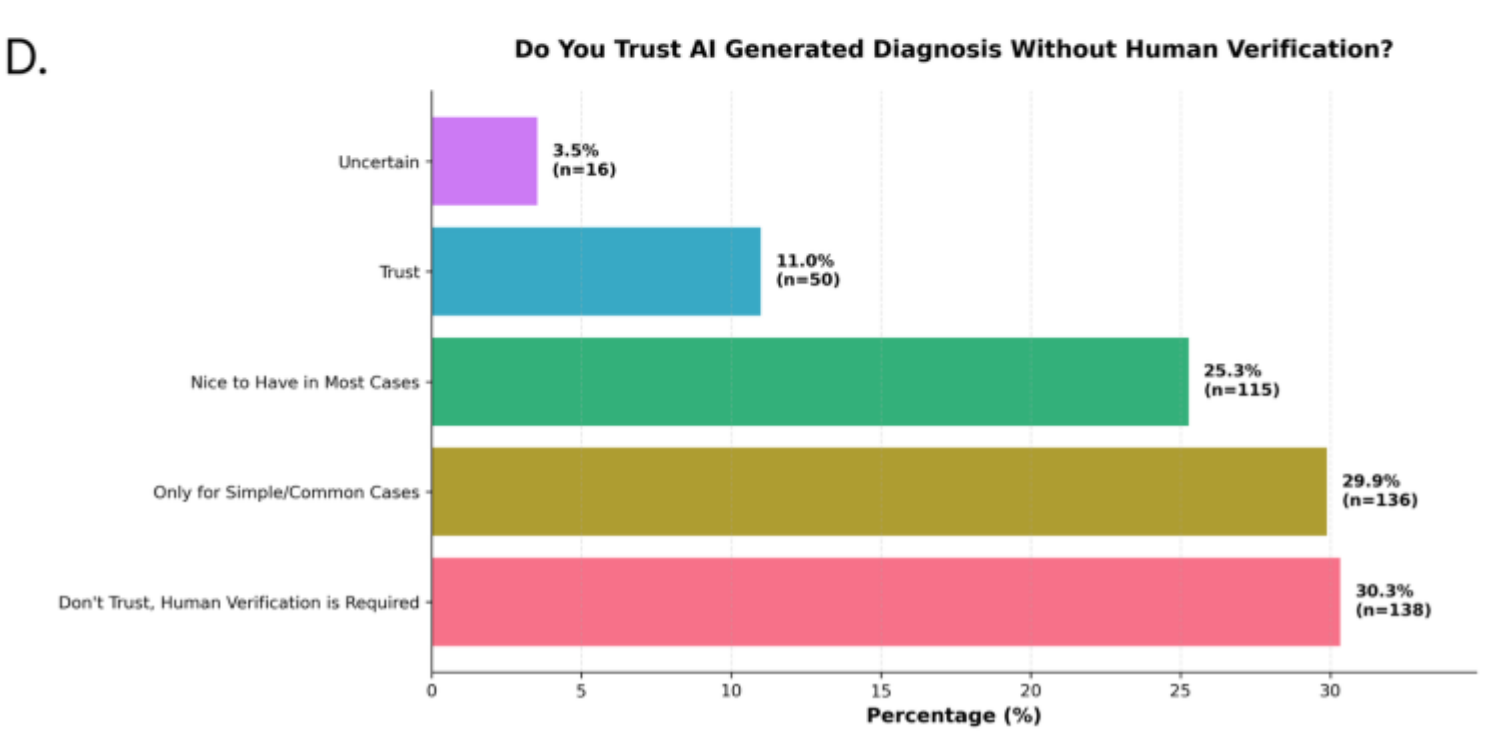

**Figure 5**. Survey results that highlight barriers to adoption and drivers for future use. Bar charts show the distribution of responses to three survey questions: (A) Biggest Obstacles to Using AI Tools in Clinics, (B) Factors That Would Promote Your AI Usage, and (C) Should AI Tool Usage Be Regulated by Veterinary Authorities? Percentages for each category are displayed on the bars, along with the number of respondents (n) for that category.

A.

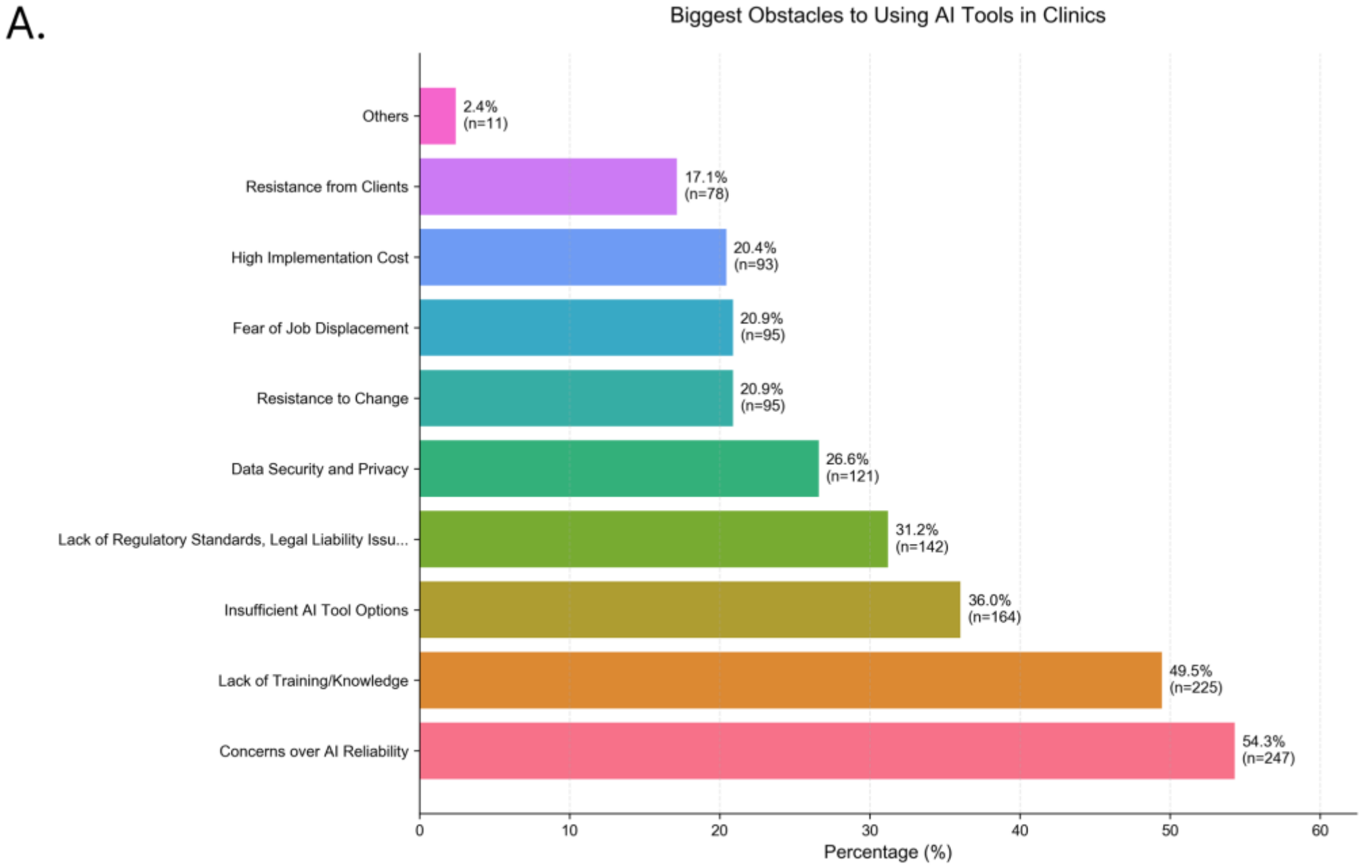

B.

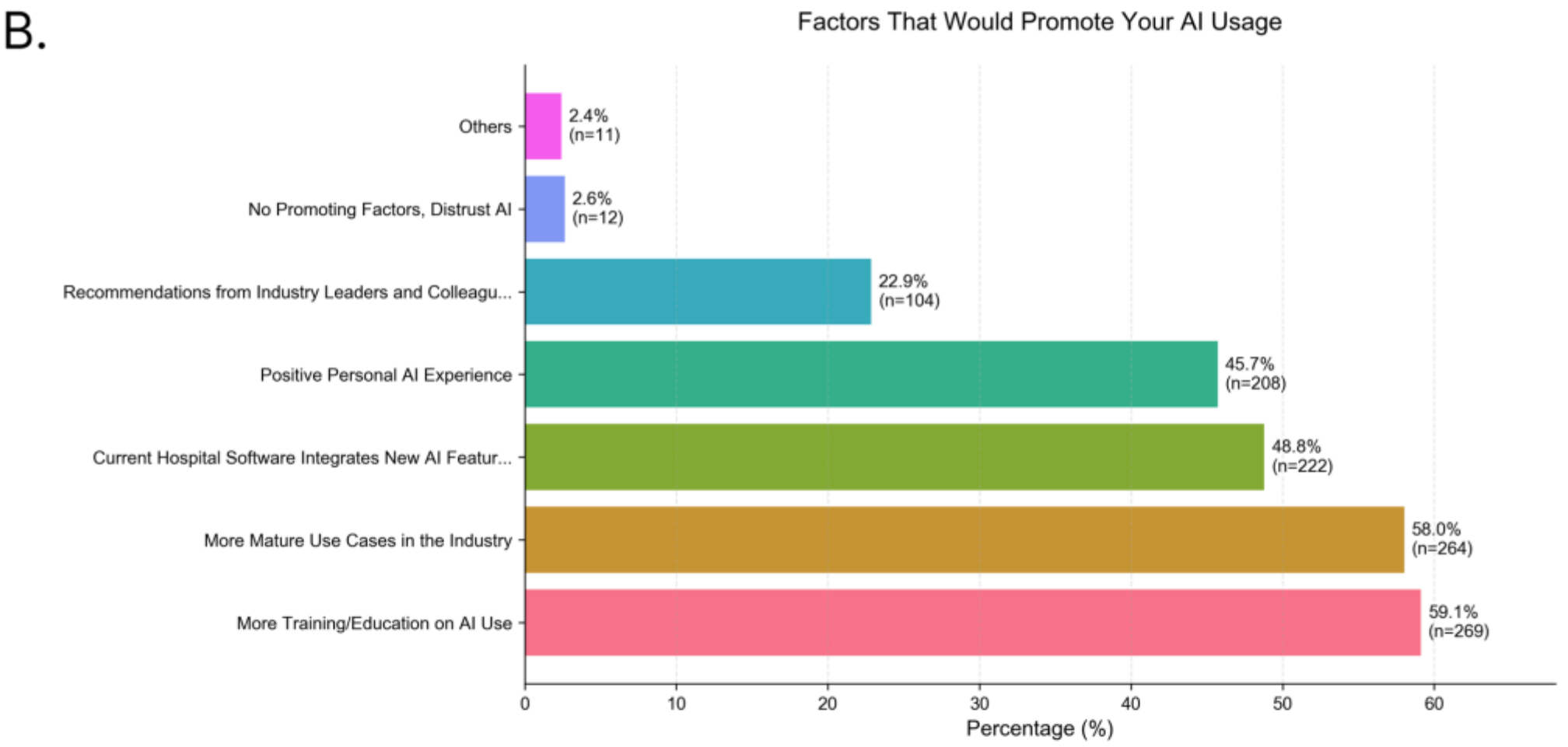

C.

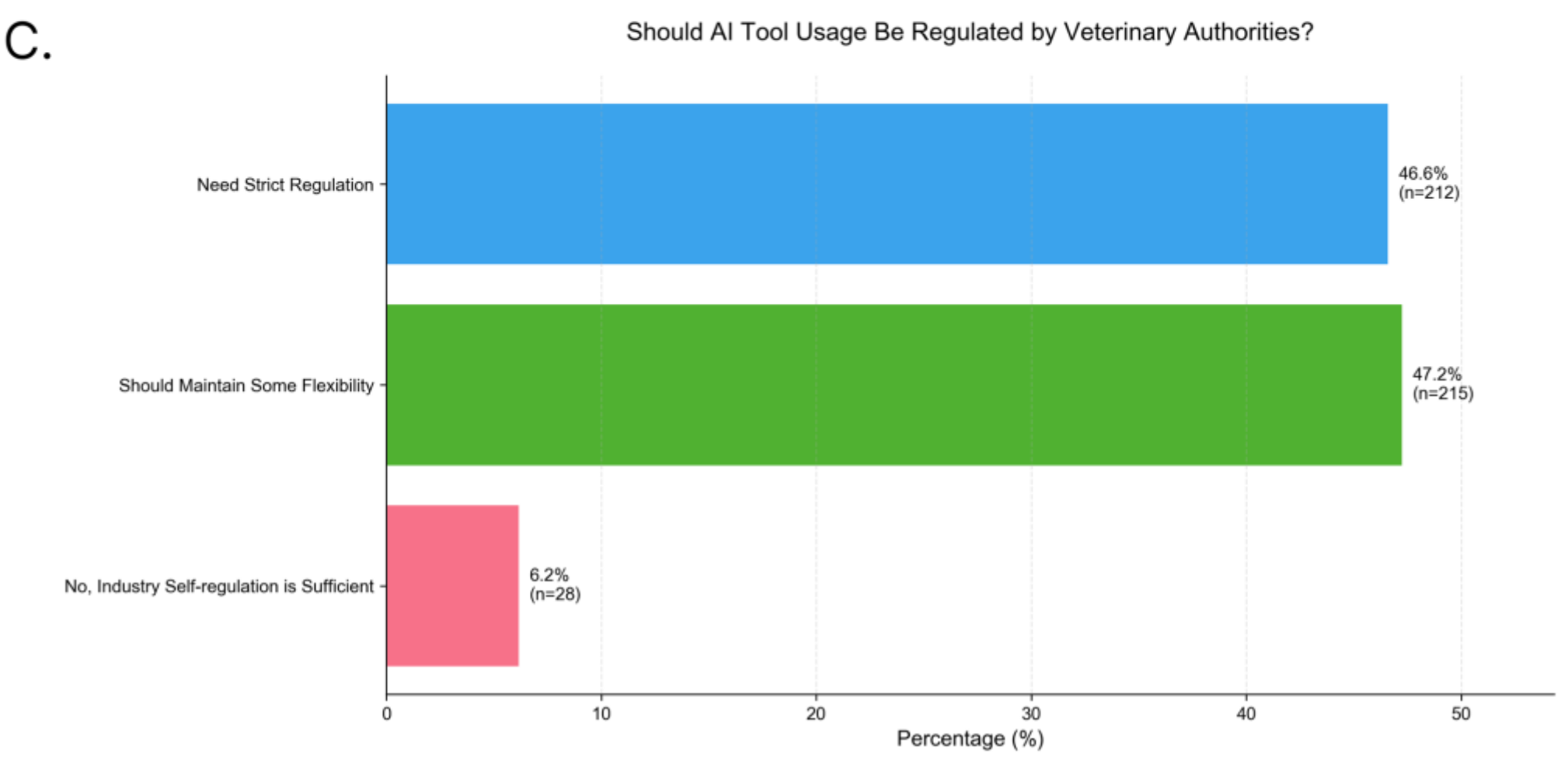

**<u>Supplemental material:</u>**

## Full questionnaire - Original version (Chinese)

### I. 基本信息

1. 您的年龄?
   - a. 00 后
   - b. 90 后
   - c. 80 后
   - d. 70 后
   - e. 60 后
2. 您在兽医学领域的主要工作是?
   - a. 兽医
   - b. 兽医技师/助理/护士
   - c. 兽医学生
   - d. 诊所管理者
   - e. 诊所前台
   - f. 其他
3. 您从事兽医学领域的工作时长?
   - a. < 1 年
   - b. 1-5 年
   - c. 6-10 年
   - d. > 10 年
4. 您所在的机构类型?
   - a. 宠物医院或诊所
   - b. 大动物诊疗机构
   - c. 异宠诊疗机构
   - d. 疫苗/宠物食品 /保健品等公司
   - e. 高校或科研院所
   - f. 政府

### II. AI 在个人和医疗上的使用

5. 您对 AI 在兽医学中的应用有多熟悉，如 ChatGPT, Deepseek?
   - a. 非常熟悉
   - b. 比较熟悉
   - c. 不太熟悉

d. 完全不熟悉
6. 您是否在兽医实践中使用过 AI 工具？
    a. 是
    b. 否
7. 您在兽医实践中使用 AI 工具的频率？
    a. 每天
    b. 每周
    c. 每月
    d. 很少使用，或从未用过
8. 以下涉及 AI 的工具或领域您了解或使用过哪些？(可多选)
    a. 放射影像分析（如 AI 辅助 X 光片解读）
    b. 电子健康记录（EHR）以及医院管理工作的自动化
    c. 语音对文字的自动转化
    d. AI 辅助的疾病诊疗和诊断
    e. 宠主的沟通和教育
    f. 对患者病史的分析
    g. 治疗方案的设计
    h. 临床上的分诊
    i. 市场营销和推广
    j. 处方药的计算和使用
    k. 虚拟多功能兽医助手
    l. 其他
9. **您是否在个人的生活中使用过** AI 工具？(可多选)
    a. Deepseek
    b. 豆包
    c. Kimi Chat
    d. 秘塔 AI
    e. ChatGLM/智谱清言
    f. 宠智灵科技
    g. 谛宝扁鹊
    h. 微宠医
    i. 贝塔猫
    j. ChatGPT, Gemini, Claude, Llama, Grok 等大语言模型
    k. iCHONG AI

l. 其他
m. 无使用经验

10. 您对 AI 工具的使用体验
    a. **试过，但不适合我/不满意效果**
    b. **尝试了一些，准备继续探索**
    c. **在日常生活中经常使用**

## III. **对** AI 应用的态度和看法

11. **您对 AI 在兽医领域的应用持什么态度?**
    a. **非常怀疑**
    b. **有一定质疑**
    c. **中立**
    d. **有一定信心**
    e. **非常有信心**
12. 您认为 AI 为兽医领域带来的主要好处是？(可多选)
    a. 提高效率，节省时间
    b. 减少病例输入及管理的负担
    c. 减少行政相关工作
    d. 提高诊疗的准确率
    e. 改善患者的预后情况
    f. 不认为有任何好处
    g. 其他
13. 以下哪个领域您认为会最大程度地享受 AI 带来的红利？(可多选)
    a. 宠主的沟通和教育
    b. 患者病例的填写
    c. 诊疗前的初诊和分诊
    d. 后续的返诊
    e. 治疗方案的安排
    f. 远程医疗
    g. 诊断检查，如血检、病理切片分析等
    h. 鉴别诊断
14. 您认为在诊所中使用 AI 工具最大的障碍是什么？(可多选)
    a. 使用成本高
    b. 缺乏培训/知识
    c. 对变革/变化的抵触

d. 对 AI 可靠性的担忧
e. AI 工具可选性不足
f. 缺乏监管标准，涉及的法律责任等
g. 担心被 AI 取代，工作机会流失
h. 内部数据的安全性没有保障
i. 宠主/畜主的抵触
j. 其他

15. 您认为以下哪些因素会促进您对 AI 的使用？(可多选)
    a. 行业出现较多比较成熟的使用案例
    b. 更多的培训机会去增加对 AI 使用的了解
    c. 个人对 AI 使用的正面经历
    d. 诊所当下使用的软件整合了新的 AI 功能
    e. 行业中其他领军人以及同事的推荐
    f. 没有可以促进的因素，不信任 AI
    g. 其他
16. 您认为对兽医从业者对 AI 方面的培训有多重要？
    a. 非常重要
    b. 比较重要
    c. 不重要
    d. 不确定
17. 您是否信任未经人工核查的 AI 初步诊断？
    a. 信任
    b. 多数情况表现良好
    c. 仅信任简单或常见病例
    d. 不信任，AI 仅仅是辅助工具，必须进行人工复核
    e. 不确定
18. 您认为畜主/宠主会接受 AI 辅助的诊断/治疗么？
    a. 大多数接受
    b. 部分接受，部分犹豫
    c. 大部分倾向纯人工诊疗
    d. 不确定
19. 您认为结合 AI 工具会提高诊所的竞争力么？
    a. 是
    b. 否

20. 您在短期内会考虑进一步扩大诊所内 AI 的使用频率和范围么？
    a. 不考虑
    b. 不确定
    c. 会考虑，但是会谨慎使用
21. 您认为 AI 工具的使用是否应由兽医监管机构统一规范？
    a. 需要严格监管
    b. 应保留一定灵活性
    c. 否，行业自调即可

## IV. 开放性问题

22. 请描述您在兽医实践中使用 AI 的积极或者消极的经历。
23. 您对 AI 在动物医疗的伦理问题有何担忧。
24. 您希望未来的 AI 兽医工具具备哪些功能。
25. 您认为未来十年 AI 将如何改变兽医的角色。
26. 关于 AI 在兽医学中的应用，您还有其他意见或建议么？

## **Full questionnaire – Translated version (English)**

I. Demographic Information

1. Your Age:

    a. Post-00s (Born after 2000)
    b. Post-90s (Born after 1990)
    c. Post-80s (Born after 1980)
    d. Post-70s (Born after 1970)
    e. Post-60s (Born after 1960)

2. Your Role in Veterinary Field

    a. Veterinarian
    b. Veterinary Technician/Assistant/Nurse
    c. Veterinary Student
    d. Clinic Manager
    e. Clinic Receptionist
    f. Other

3. Your Work Experience in Veterinary Field

    a. < 1 year

b. 1-5 years
c. 6-10 years
d. > 10 years

4. Type of Institution You Work At

a. Pet Hospital/Clinic
b. Large Animal Clinic
c. Exotic Pet Clinic
d. Vaccine/Pet Food/Health Product Company
e. University/Research Institute
f. Government

II. AI Use in Personal and Medical Applications

5. Your Familiarity with AI Applications in Veterinary Medicine, such as ChatGPT, Deepseek?

a. Very Familiar
b. Somewhat Familiar
c. Not Very Familiar
d. Completely Unfamiliar

6. Have You Used AI Tools in Veterinary Practice?

a. Yes
b. No

7. Frequency of AI Tool Usage in Veterinary Practice

a. Daily
b. Weekly
c. Monthly
d. Rarely use

8. Which AI Tools or Fields Have You Known or Used? (Multiple Choice)

a. AI-assisted Radiology Interpretation
b. EHR & Practice Management
c. Voice-to-Text Transcription
d. AI-assisted Diagnostics
e. Pet Owner Communication and Education
f. Patient History Analysis
g. Treatment Planning
h. Triage
i. Marketing
j. Prescription & Dosage Calculation

k. Virtual Assistant
l. Other

9. Have you used AI tools in your personal life? (Multiple Choice)

a. Deepseek
b. DouBao
c. Kimi Chat
d. Mita AI
e. ChatGLM/Zhipu AI
f. Chongzhiling
g. Dibaobianque
h. Mini Vet
i. Beta Cat
j. ChatGPT, Gemini, Claude, Llama, Grok and other LLMs
k. iCHONG AI
l. Other
m. Never Used

10. Your Experience with AI Tools

a. Tried, but unsatisfied with results
b. Tried some, plan to continue
c. Positive experience, frequently used in daily life

III. Attitudes and Perspectives on AI Applications

11. Your Attitude Toward AI Applications in Veterinary Field

a. Very Skeptical
b. Somewhat Skeptical
c. Neutral
d. Somewhat Confident
e. Very Confident

12. Main Benefits AI Brings to Veterinary Field (Multiple Choice)

a. Improve Efficiency, Save Time
b. Reduce EHR related Workload
c. Reduce Administrative Workload
d. Improve Diagnostic Accuracy
e. Improve Prognosis
f. No Benefits Perceived
g. Others

13. Which Field Will Benefit Most from AI? (Multiple Choice)

    a. Client Communication and Education
    b. Medical Record Writing
    c. Clinical Triage
    d. Follow-up Visits
    e. Treatment Plan
    f. Telemedicine
    g. Diagnostics, such as blood tests, pathological slide analysis, etc.
    h. Differential Diagnosis

14. Biggest Obstacles to Using AI Tools in Clinics (Multiple Choice)

    a. High Implementation Cost
    b. Lack of Training/Knowledge
    c. Resistance to Change
    d. Concerns about AI Reliability
    e. Insufficient AI Tool Options
    f. Lack of Regulatory Standards, Legal Liability Issues
    g. Fear of Job Displacement
    h. Data Security and Privacy
    i. Resistance from Clients
    j. Others

15. Factors That Would Promote Your AI Usage (Multiple Choice)

    a. More Mature Use Cases in the Industry
    b. More Training/Education on AI Use
    c. Positive Personal AI Experience
    d. Current Hospital Software Integrates New AI Features
    e. Recommendations from Industry Leaders and Colleagues
    f. No Promoting Factors, Distrust AI
    g. Others

16. Importance of AI Training for Veterinary Practitioners

    a. Very Important
    b. Somewhat Important
    c. Not Important
    d. Uncertain

17. Do You Trust AI Preliminary Diagnosis Without Human Verification?

    a. Trust
    b. Good Performance in Most Cases

c. Only Trust for Simple/Common Cases
d. Don't Trust, AI is Only Assistive, Human Verification is Required
e. Uncertain

18. Do You Think Pet Owners Will Accept AI-Assisted Diagnosis/Treatment?

a. Most Accept
b. Some Accept, Some Hesitate
c. Most Prefer Veterinarians
d. Uncertain

19. Can AI Tools Improve Medical Institution Competitiveness?

a. Yes
b. No

20. Will You Consider Expanding AI Usage in Your Institution in the Short Term?

a. Not Considering
b. Uncertain
c. Will Consider, But Use Cautiously

21. Should AI Tool Usage Be Regulated by Veterinary Authorities?

a. Need Strict Regulation
b. Should Maintain Some Flexibility
c. No, Industry Self-regulation is Sufficient

IV. Open-ended Questions

22. Please describe your positive or negative experiences using AI in veterinary practice.

23. What are your concerns about the ethical issues of AI in animal healthcare?

24. What features do you hope future AI veterinary tools will have?

25. How do you think AI will change the role of veterinarians in the next ten years?

26.Do you have any other opinions or suggestions regarding the application of AI in veterinary medicine?

# Supplementary Figures: Visualizations of Response Distributions (Questions 1–21)

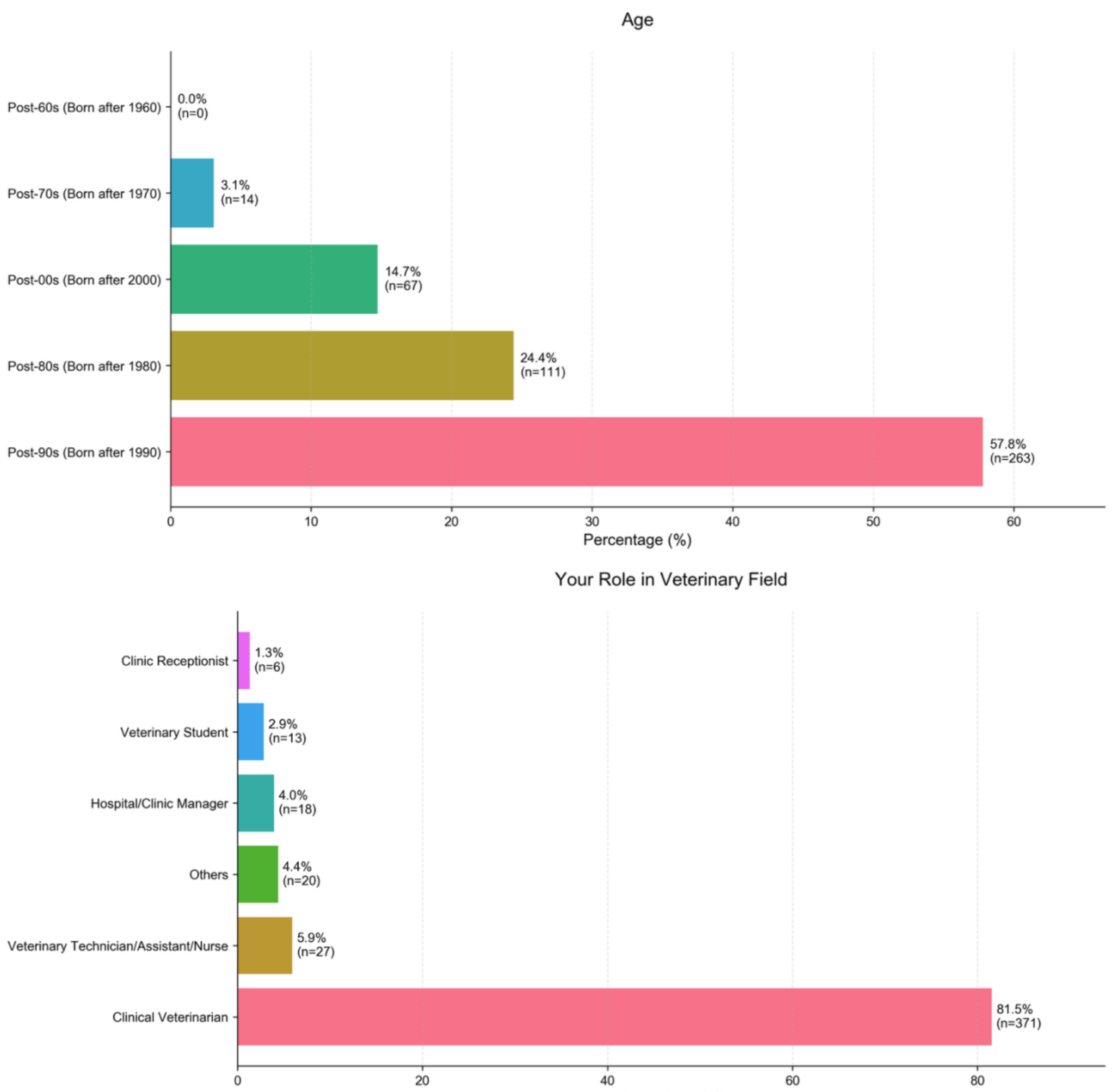

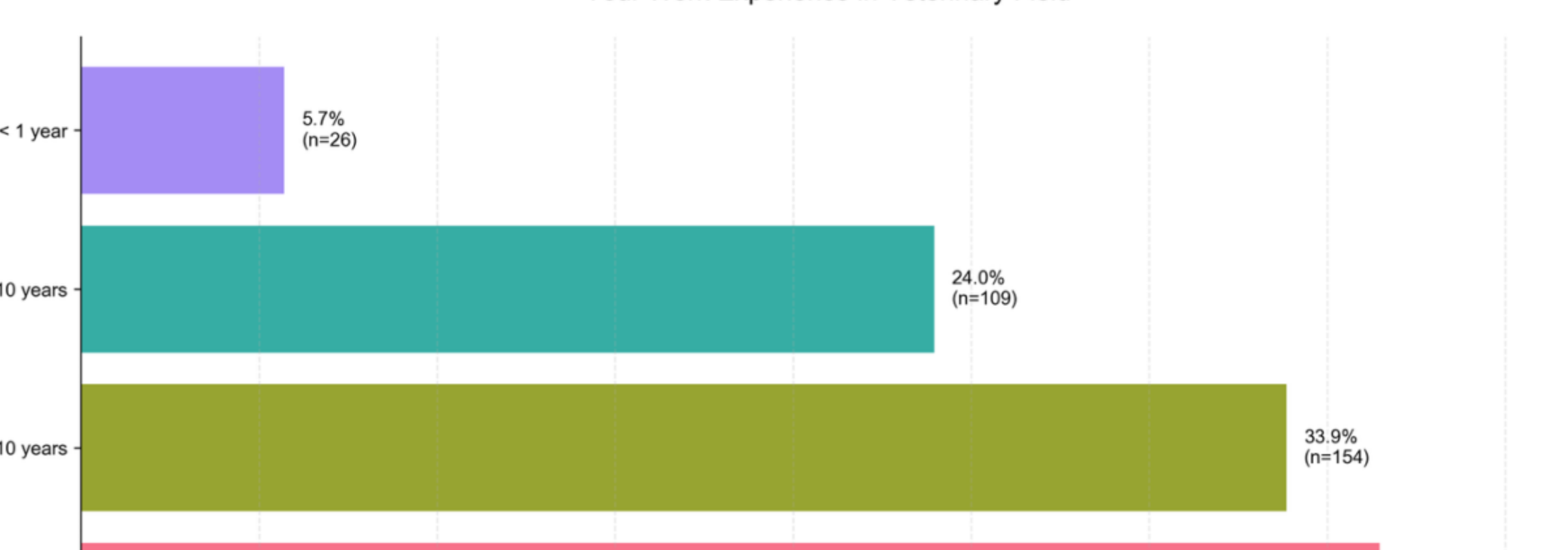
Your Work Experience in Veterinary Field
< 1 year
5.7%
(n=26)
> 10 years
24.0%
(n=109)
6-10 years
33.9%
(n=154)
1-5 years
36.5%
(n=166)
0
5
10
15
20
25
30
35
40
Percentage (%)

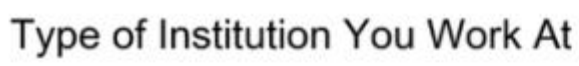
Type of Institution You Work At

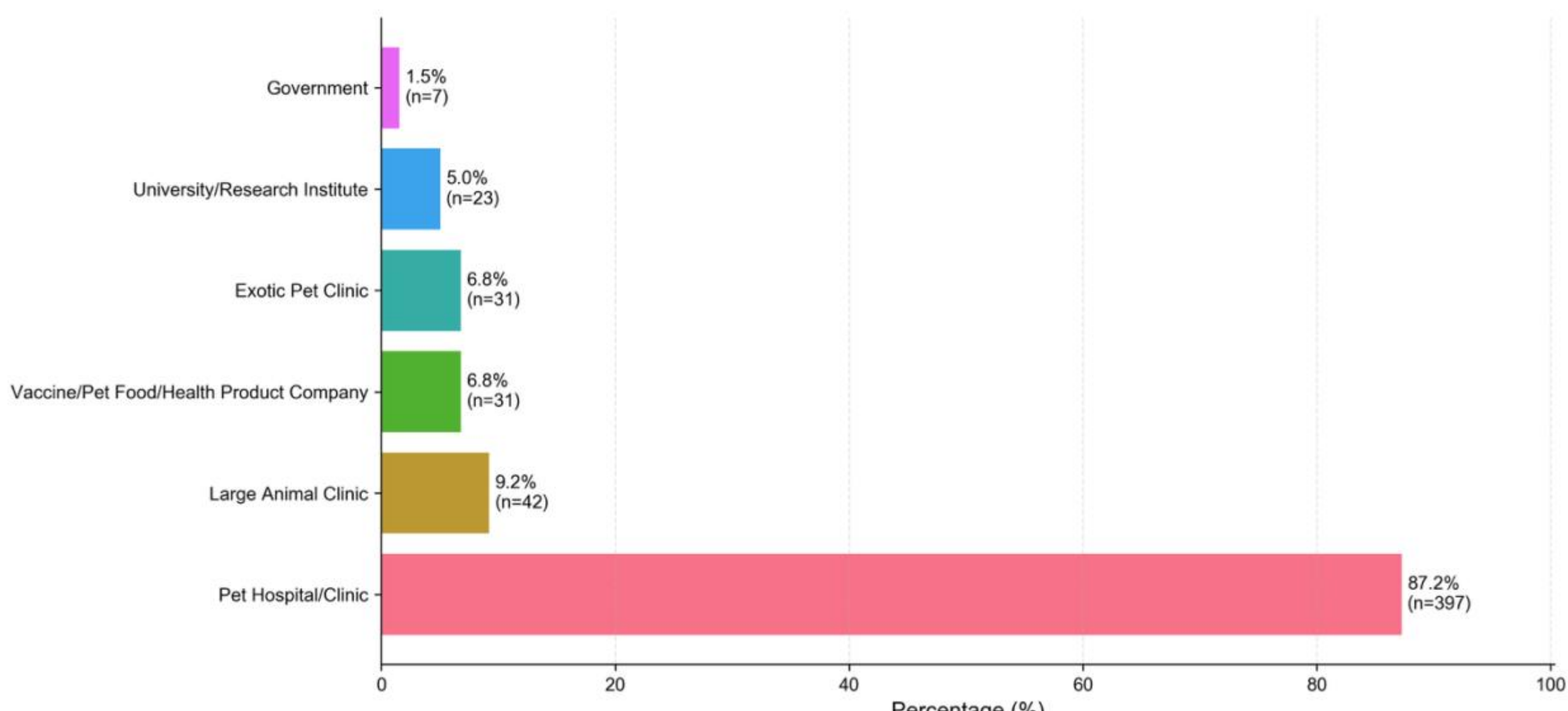
Government
1.5%
(n=7)
University/Research Institute
5.0%
(n=23)
Exotic Pet Clinic
6.8%
(n=31)
Vaccine/Pet Food/Health Product Company
6.8%
(n=31)
Large Animal Clinic
9.2%
(n=42)
Pet Hospital/Clinic
87.2%
(n=397)
0
20
40
60
80
100
Percentage (%)

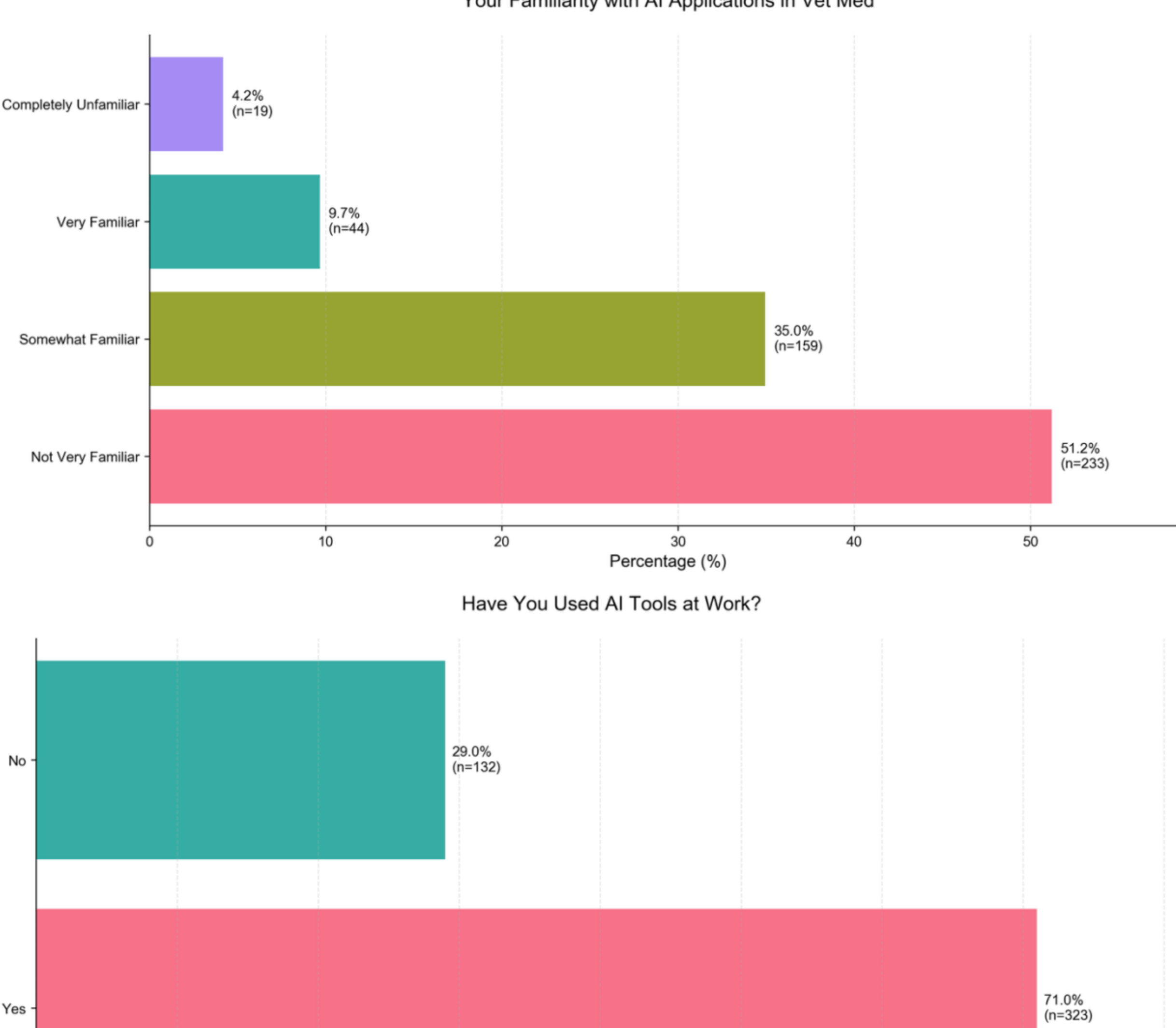
Your Familiarity with AI Applications in Vet Med
Completely Unfamiliar
4.2%
(n=19)
Very Familiar
9.7%
(n=44)
Somewhat Familiar
35.0%
(n=159)
Not Very Familiar
51.2%
(n=233)
0
10
20
30
40
50
Percentage (%)
Have You Used AI Tools at Work?
No
29.0%
(n=132)
Yes
71.0%
(n=323)
0
10
20
30
40
50
60
70
80
Percentage (%)

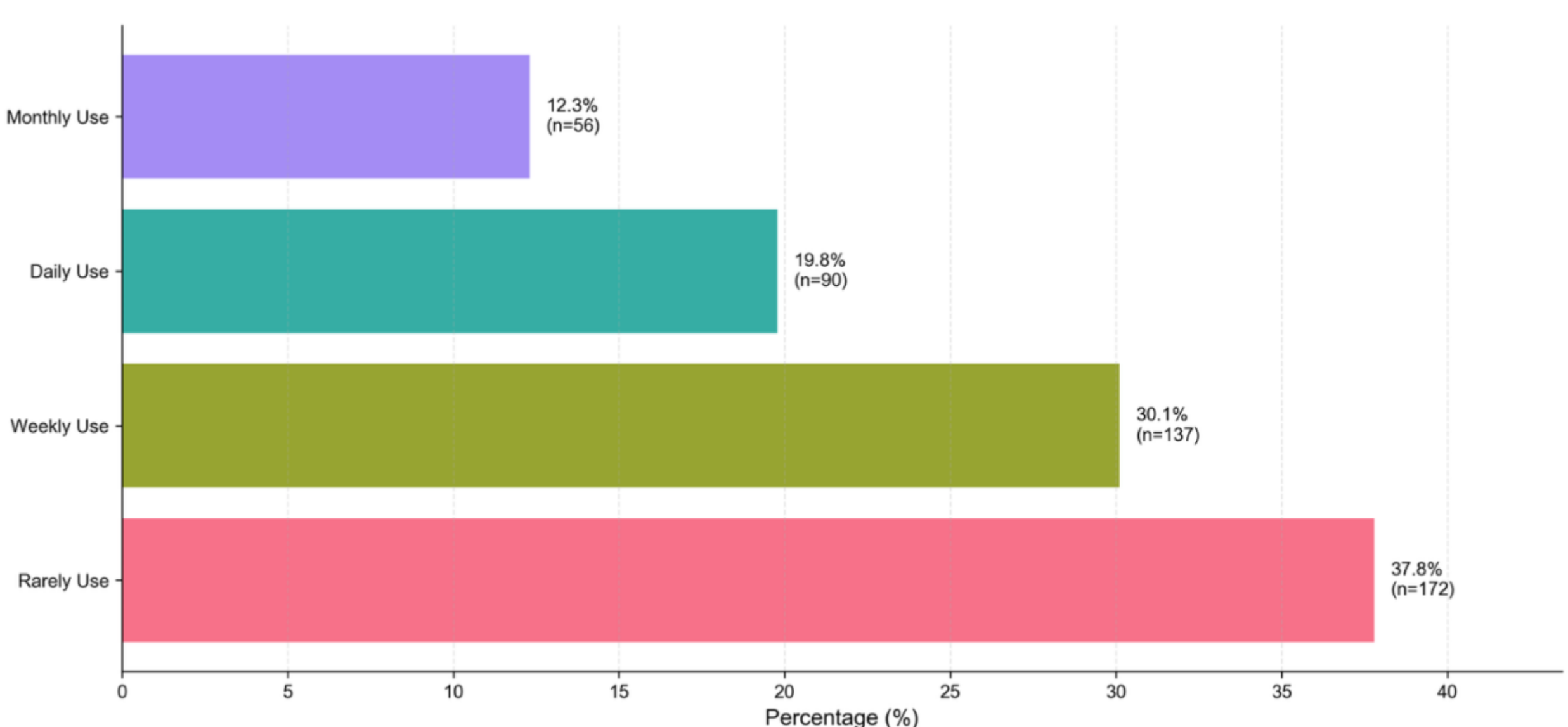
Frequency of AI Tool Usage at Work
Monthly Use
12.3%
(n=56)
Daily Use
19.8%
(n=90)
Weekly Use
30.1%
(n=137)
Rarely Use
37.8%
(n=172)
0
5
10
15
20
25
30
35
40
Percentage (%)

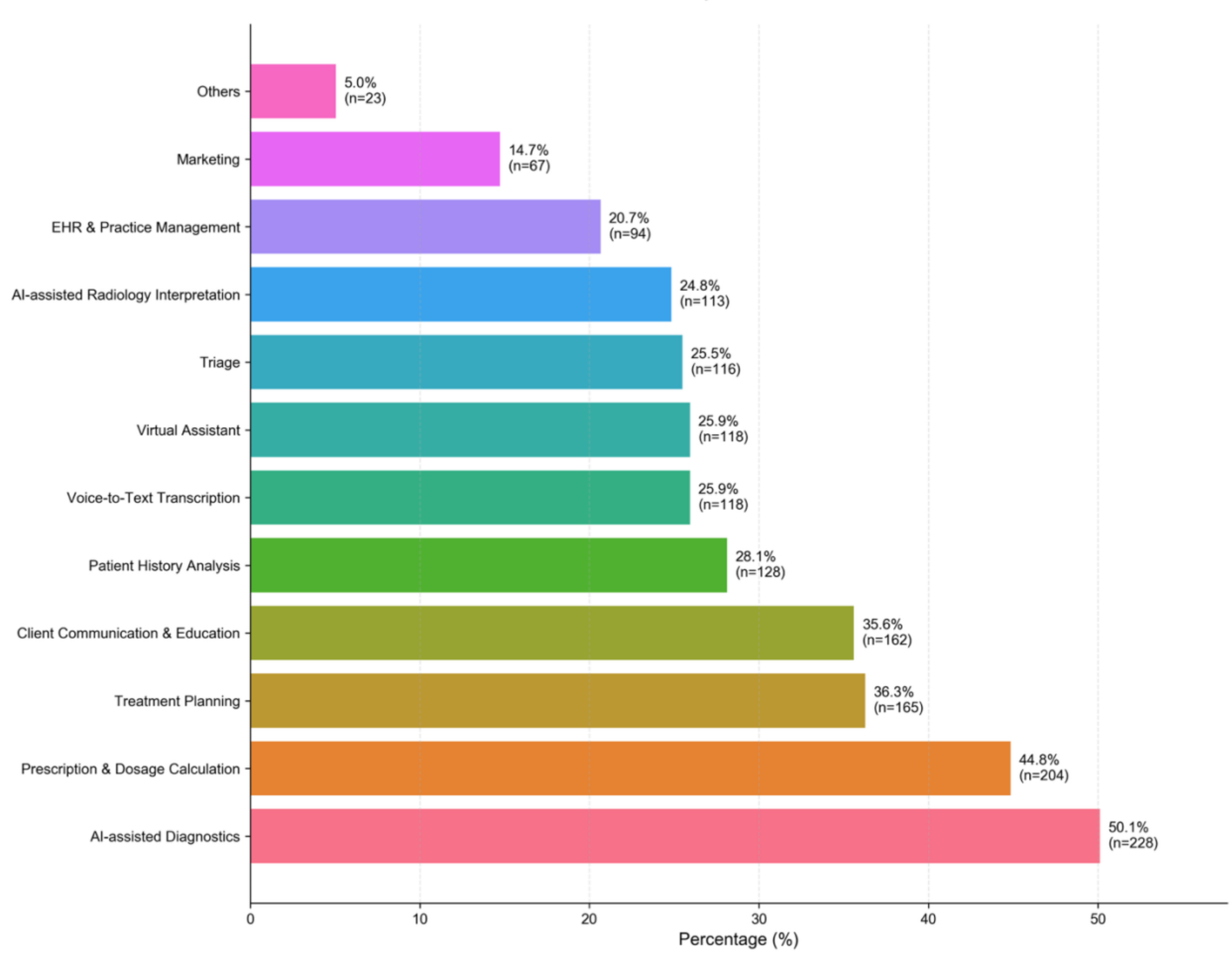
AI tools or fields you know or have used?
Others
5.0%
(n=23)
Marketing
14.7%
(n=67)
EHR & Practice Management
20.7%
(n=94)
AI-assisted Radiology Interpretation
24.8%
(n=113)
Triage
25.5%
(n=116)
Virtual Assistant
25.9%
(n=118)
Voice-to-Text Transcription
25.9%
(n=118)
Patient History Analysis
28.1%
(n=128)
Client Communication & Education
35.6%
(n=162)
Treatment Planning
36.3%
(n=165)
Prescription & Dosage Calculation
44.8%
(n=204)
AI-assisted Diagnostics
50.1%
(n=228)
0
10
20
30
40
50
Percentage (%)

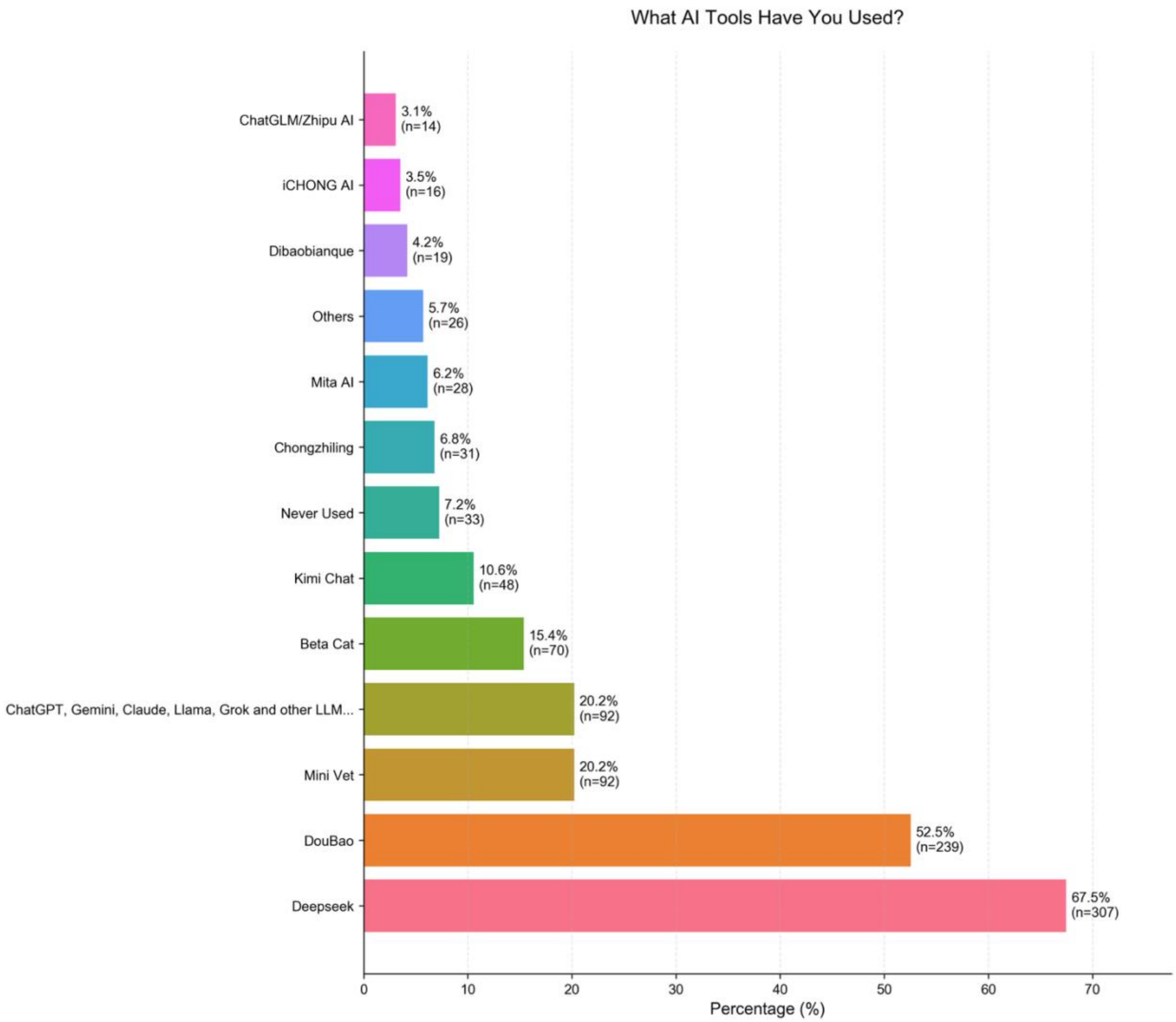

## What AI Tools Have You Used?

|  | Percentage (%) |
| --- | --- |
| ChatGLM/Zhipu AI | 3.1% (n=14) |
| iCHONG AI | 3.5% (n=16) |
| Dibaobianque | 4.2% (n=19) |
| Others | 5.7% (n=26) |
| Mita AI | 6.2% (n=28) |
| Chongzhiling | 6.8% (n=31) |
| Never Used | 7.2% (n=33) |
| Kimi Chat | 10.6% (n=48) |
| Beta Cat | 15.4% (n=70) |
| ChatGPT, Gemini, Claude, Llama, Grok and other LLM... | 20.2% (n=92) |
| Mini Vet | 20.2% (n=92) |
| DouBao | 52.5% (n=239) |
| Deepseek | 67.5% (n=307) |

Percentage (%)

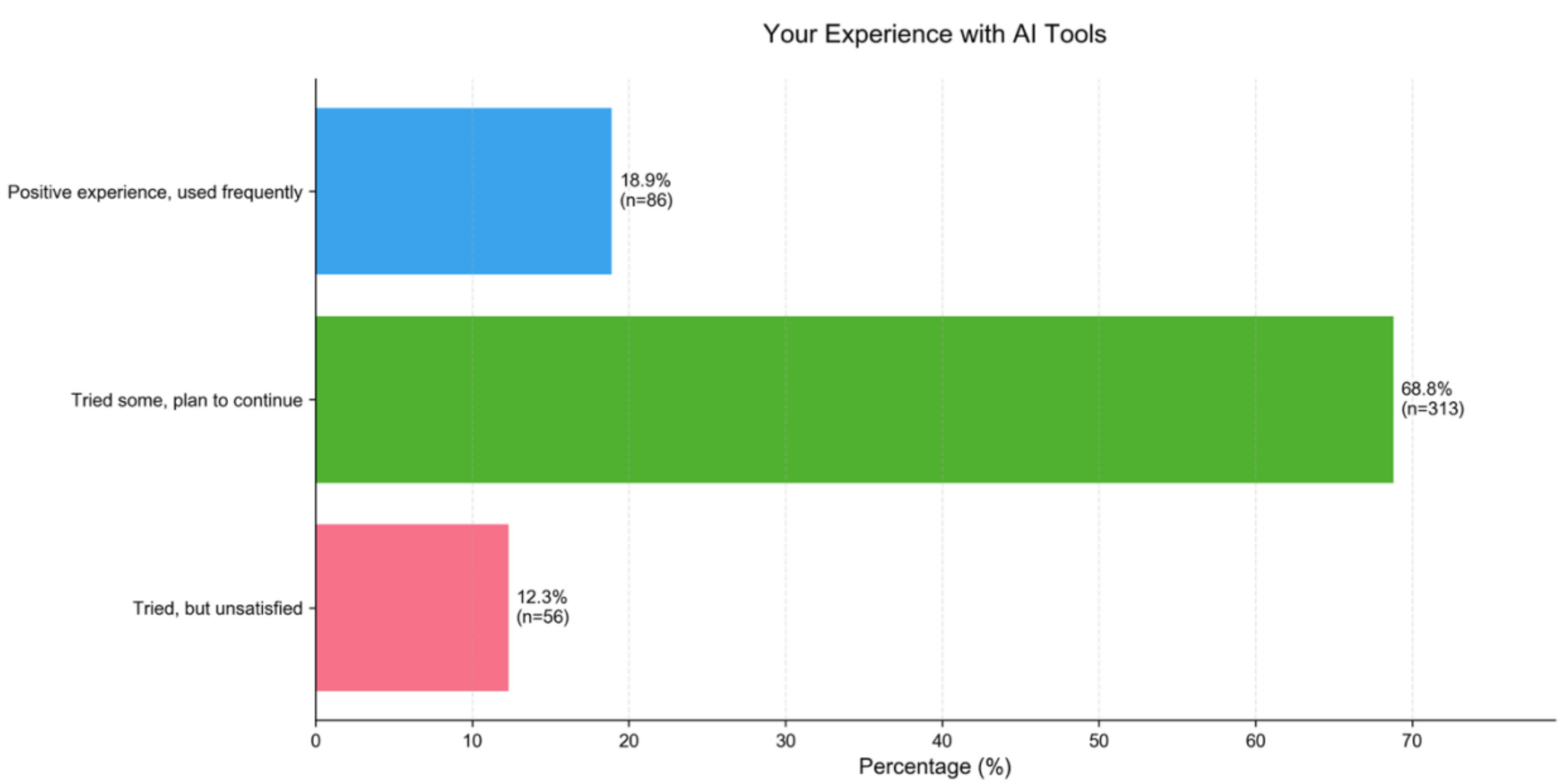

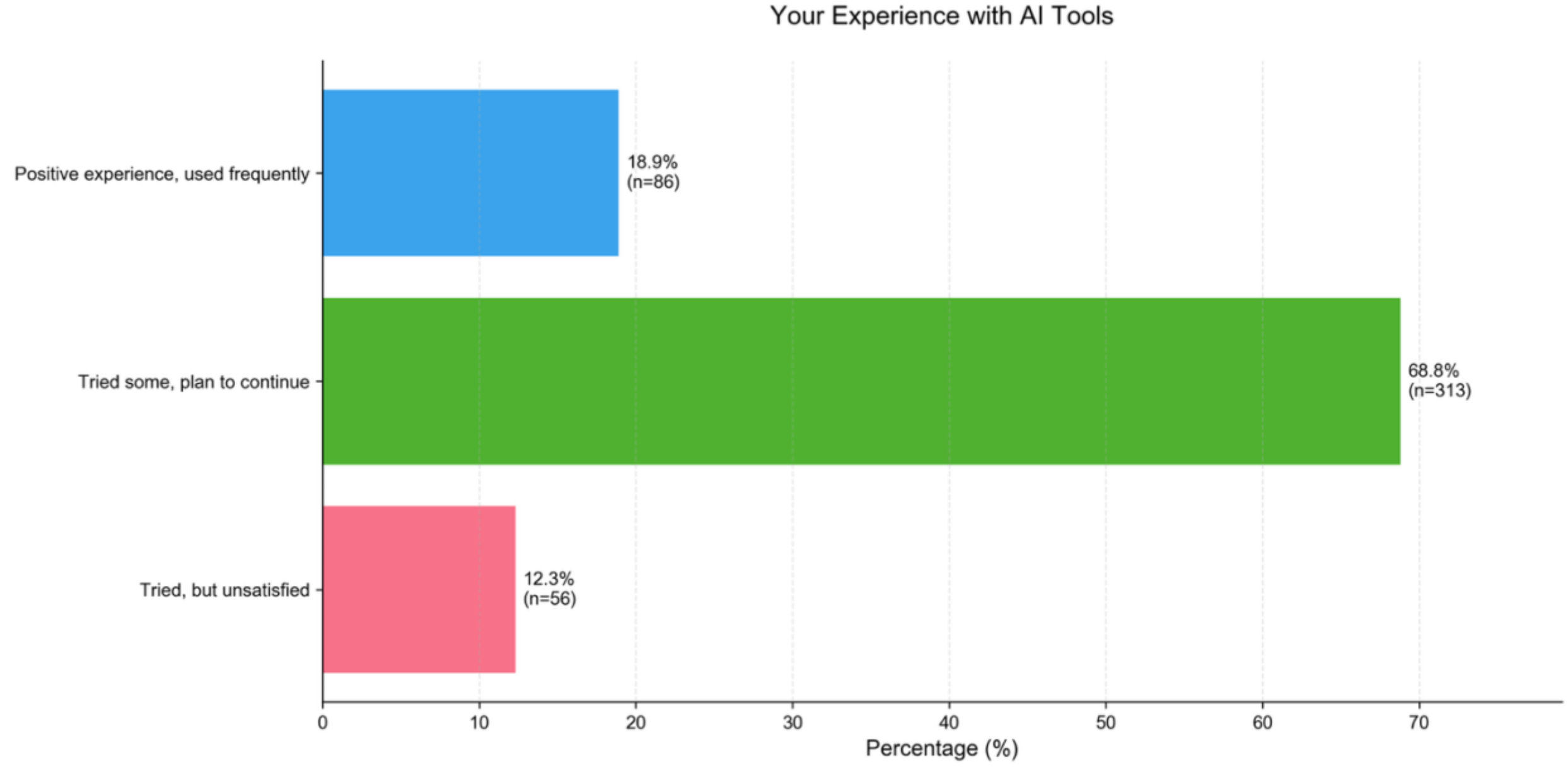

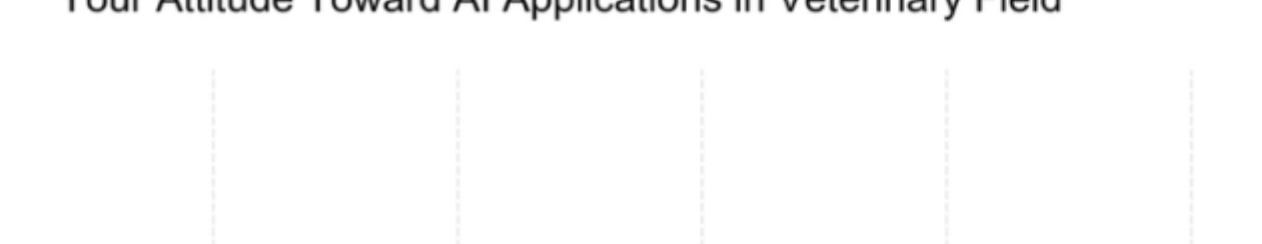
Your Attitude Toward AI Applications in Veterinary Field

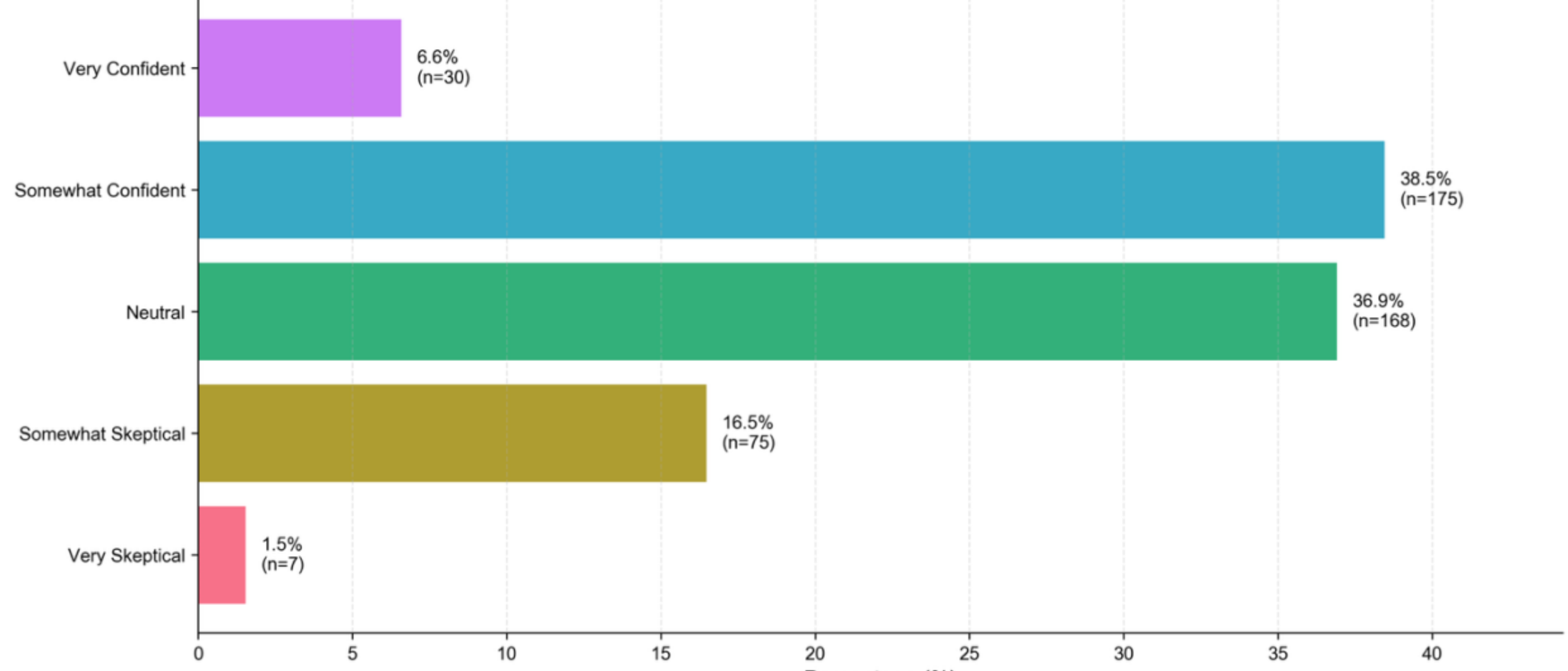
Very Confident
6.6%
(n=30)
Somewhat Confident
38.5%
(n=175)
Neutral
36.9%
(n=168)
Somewhat Skeptical
16.5%
(n=75)
Very Skeptical
1.5%
(n=7)
0
5
10
15
20
25
30
35
40
Percentage (%)

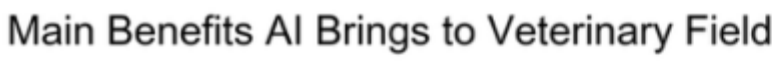
Main Benefits AI Brings to Veterinary Field

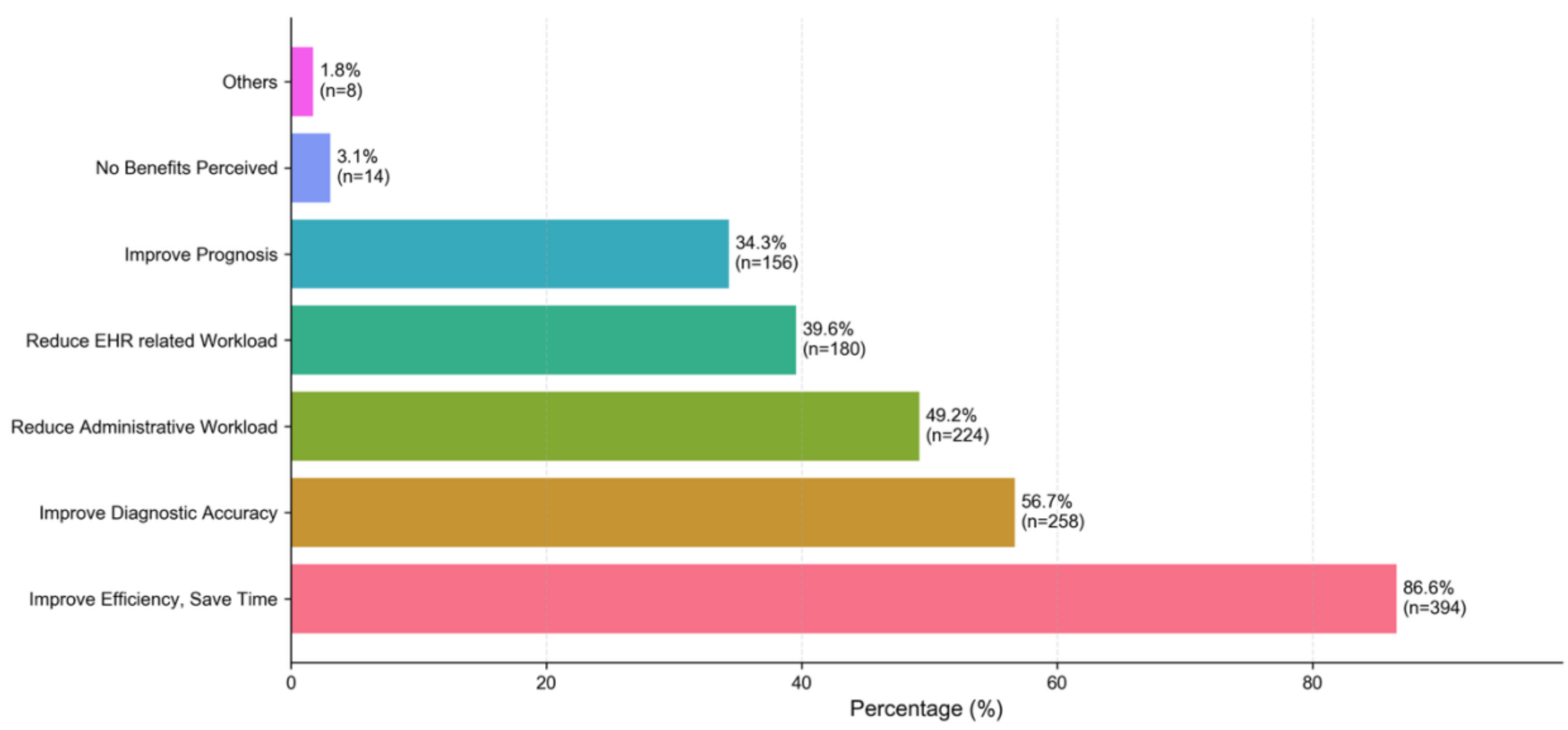
Others
1.8%
(n=8)
No Benefits Perceived
3.1%
(n=14)
Improve Prognosis
34.3%
(n=156)
Reduce EHR related Workload
39.6%
(n=180)
Reduce Administrative Workload
49.2%
(n=224)
Improve Diagnostic Accuracy
56.7%
(n=258)
Improve Efficiency, Save Time
86.6%
(n=394)
0
20
40
60
80
Percentage (%)

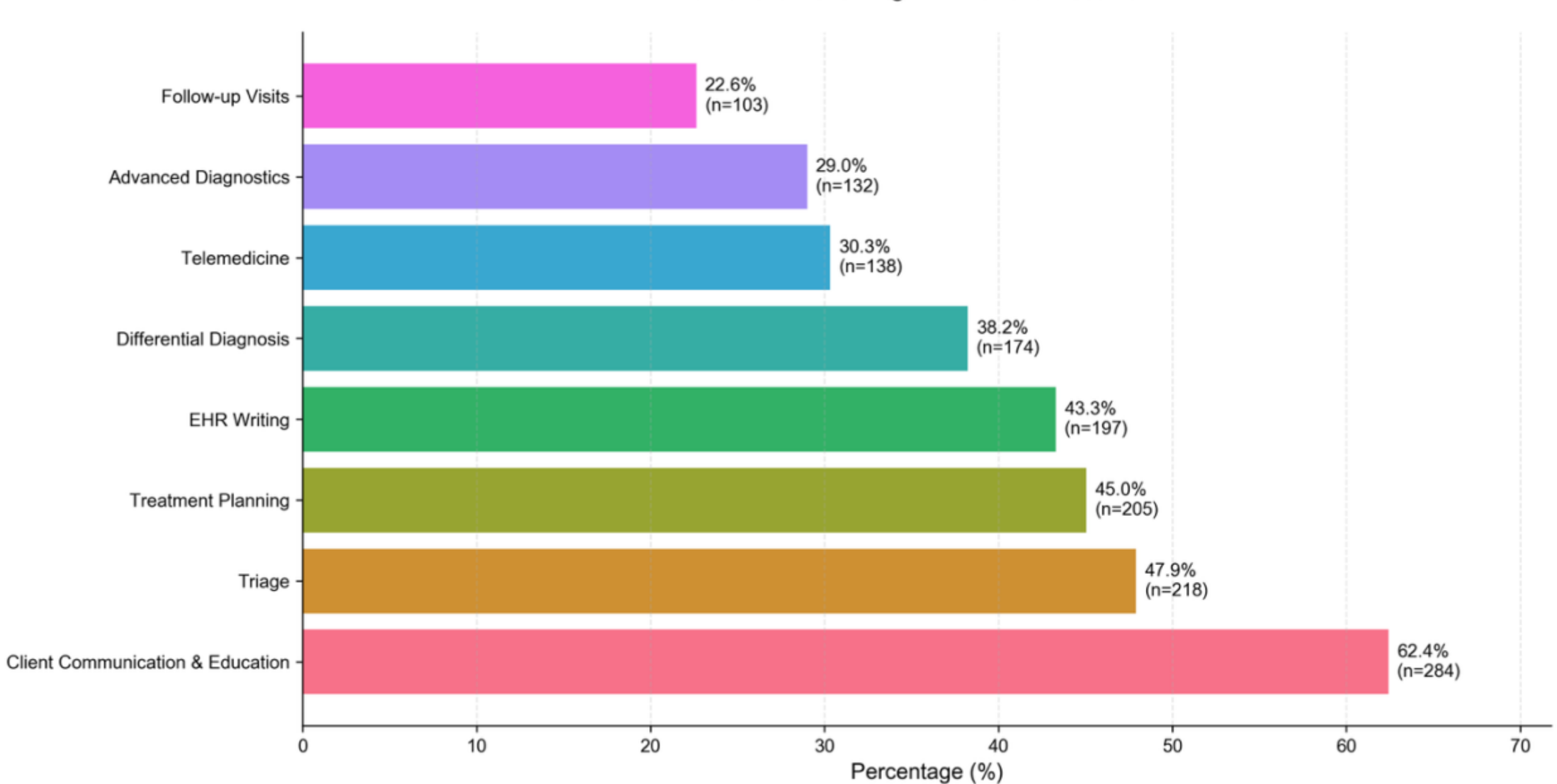
Which of the Following Will Benefit Most from AI?
Follow-up Visits
Advanced Diagnostics
Telemedicine
Differential Diagnosis
EHR Writing
Treatment Planning
Triage
Client Communication & Education
22.6% (n=103)
29.0% (n=132)
30.3% (n=138)
38.2% (n=174)
43.3% (n=197)
45.0% (n=205)
47.9% (n=218)
62.4% (n=284)
0
10
20
30
40
50
60
70
Percentage (%)

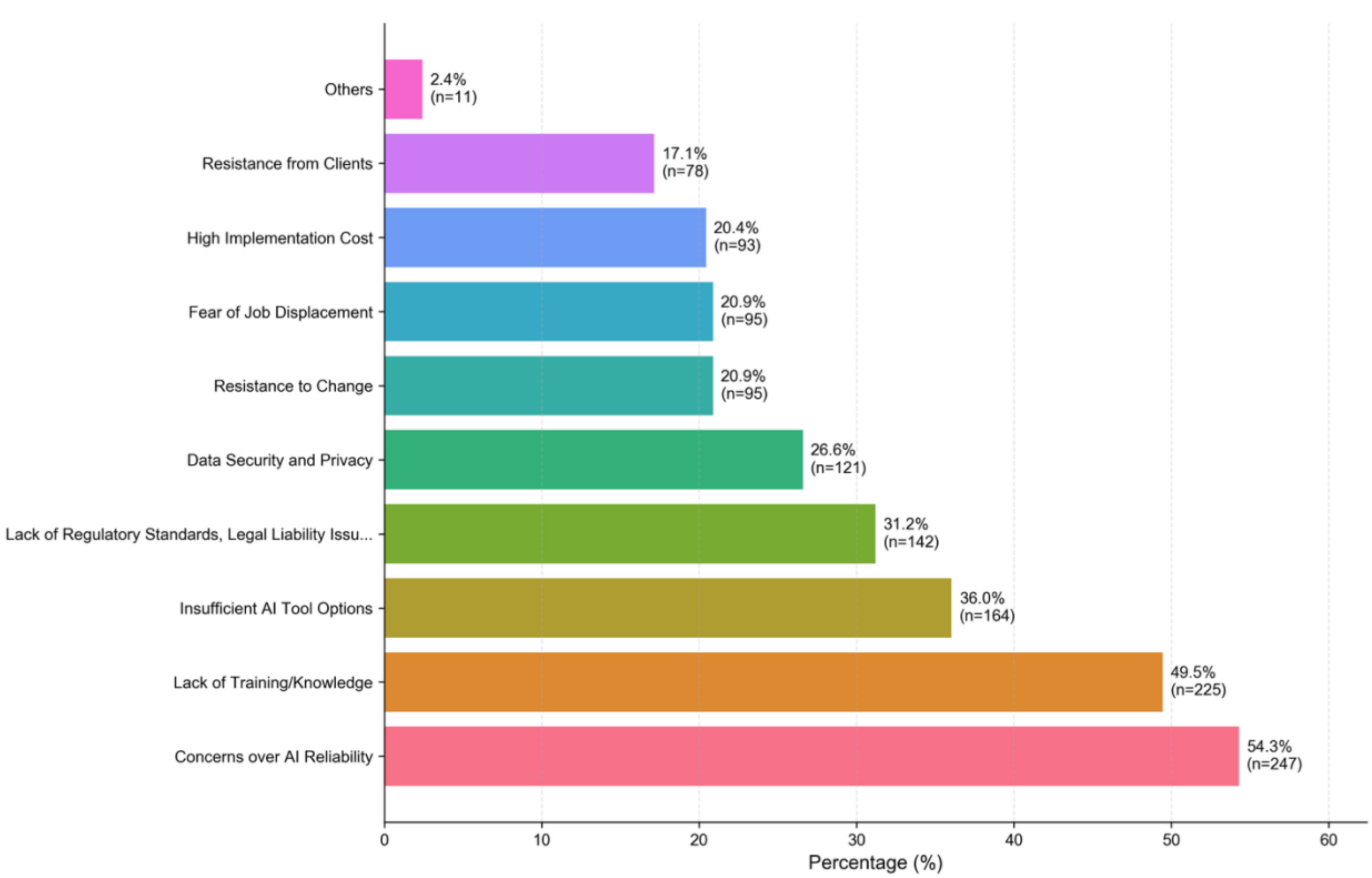
Biggest Obstacles to Using AI Tools in Clinics
Others
Resistance from Clients
High Implementation Cost
Fear of Job Displacement
Resistance to Change
Data Security and Privacy
Lack of Regulatory Standards, Legal Liability Issu...
Insufficient AI Tool Options
Lack of Training/Knowledge
Concerns over AI Reliability
2.4% (n=11)
17.1% (n=78)
20.4% (n=93)
20.9% (n=95)
20.9% (n=95)
26.6% (n=121)
31.2% (n=142)
36.0% (n=164)
49.5% (n=225)
54.3% (n=247)
0
10
20
30
40
50
60
Percentage (%)

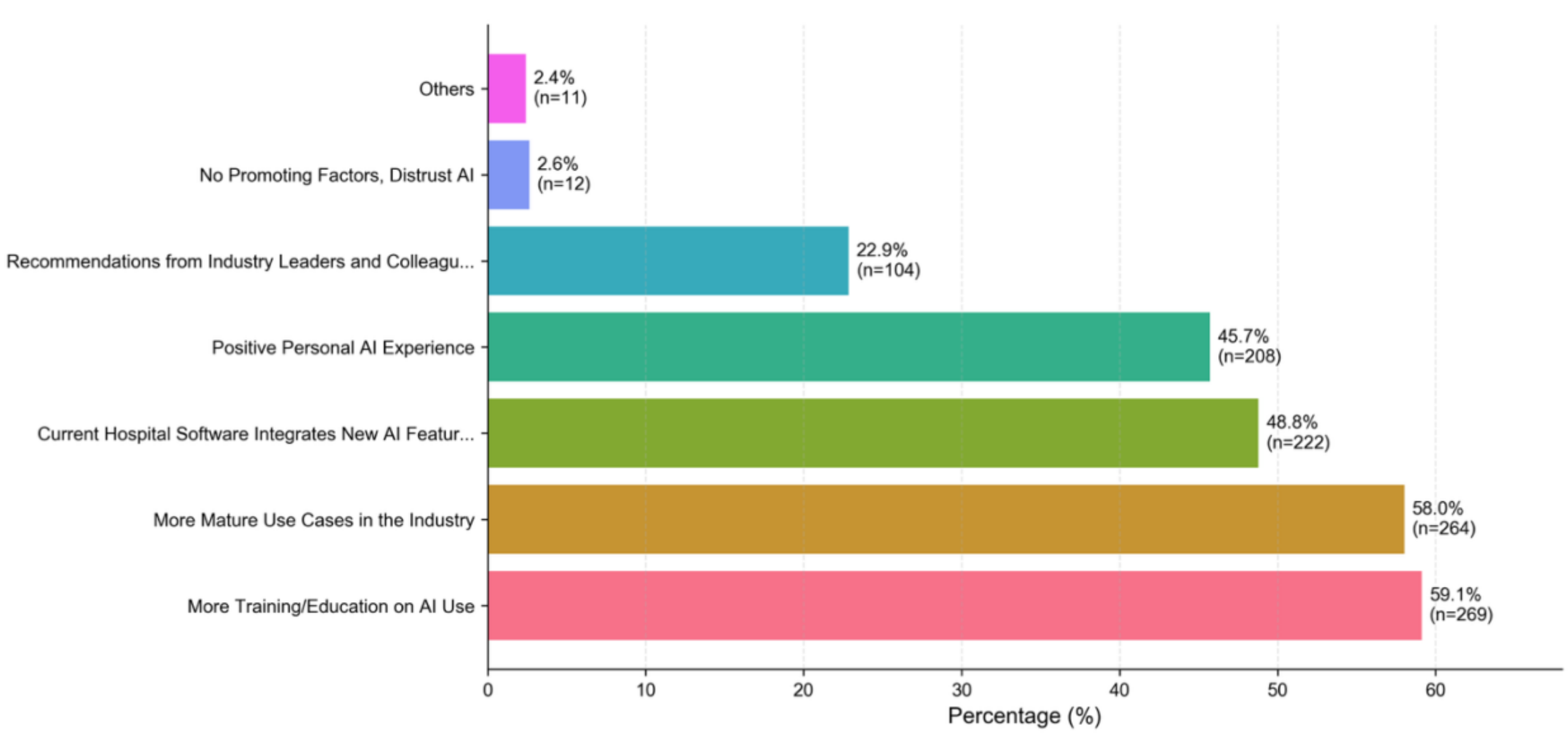

Factors That Would Promote Your AI Usage
Others
2.4%
(n=11)
No Promoting Factors, Distrust AI
2.6%
(n=12)
Recommendations from Industry Leaders and Colleagu...
22.9%
(n=104)
Positive Personal AI Experience
45.7%
(n=208)
Current Hospital Software Integrates New AI Featur...
48.8%
(n=222)
More Mature Use Cases in the Industry
58.0%
(n=264)
More Training/Education on AI Use
59.1%
(n=269)
0
10
20
30
40
50
60
Percentage (%)

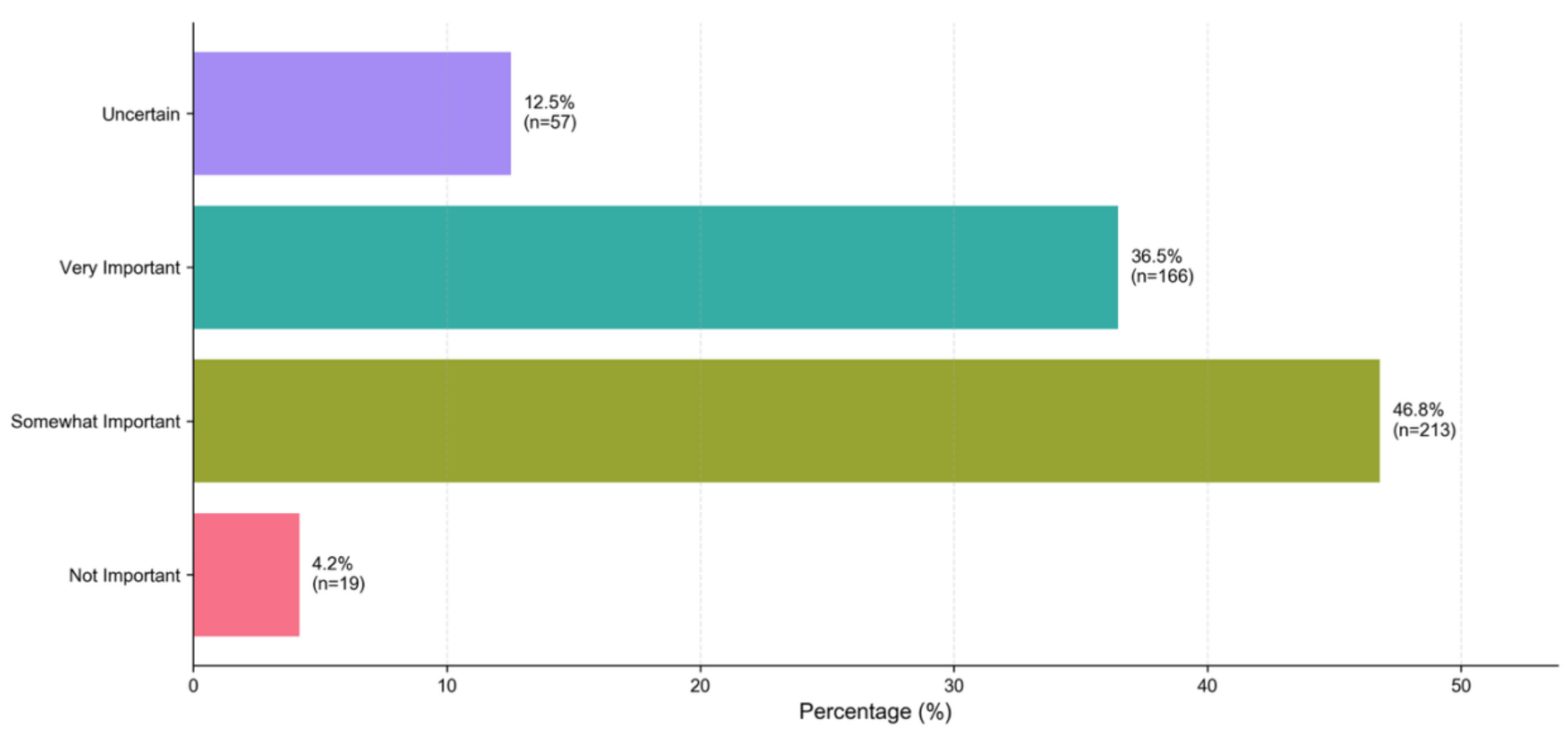

Importance of AI Training for Veterinary Practitioners
Uncertain
12.5%
(n=57)
Very Important
36.5%
(n=166)
Somewhat Important
46.8%
(n=213)
Not Important
4.2%
(n=19)
0
10
20
30
40
50
Percentage (%)

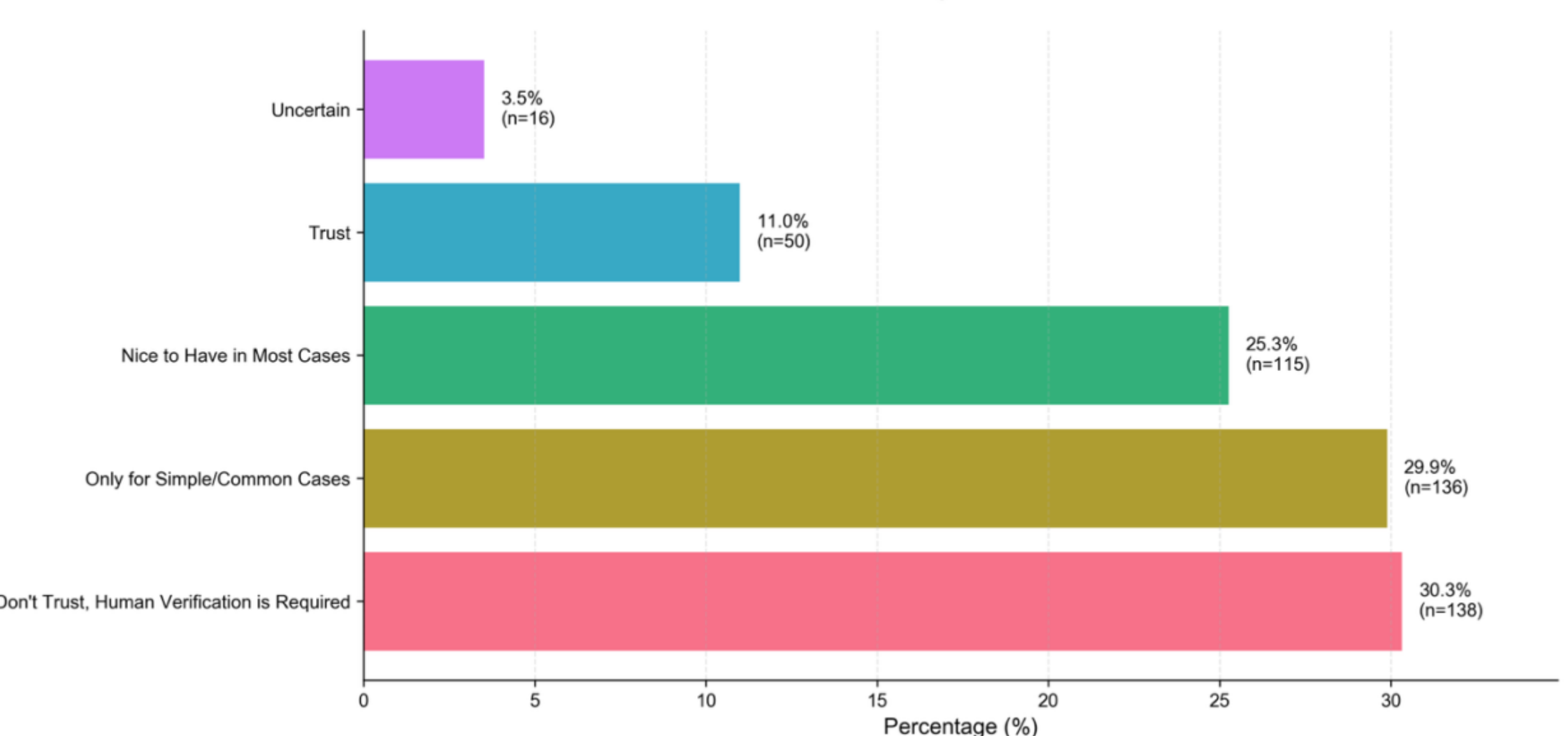
Do You Trust AI Generated Diagnosis Without Human Verification?
Uncertain
Trust
Nice to Have in Most Cases
Only for Simple/Common Cases
Don't Trust, Human Verification is Required
3.5% (n=16)
11.0% (n=50)
25.3% (n=115)
29.9% (n=136)
30.3% (n=138)
0
5
10
15
20
25
30
Percentage (%)

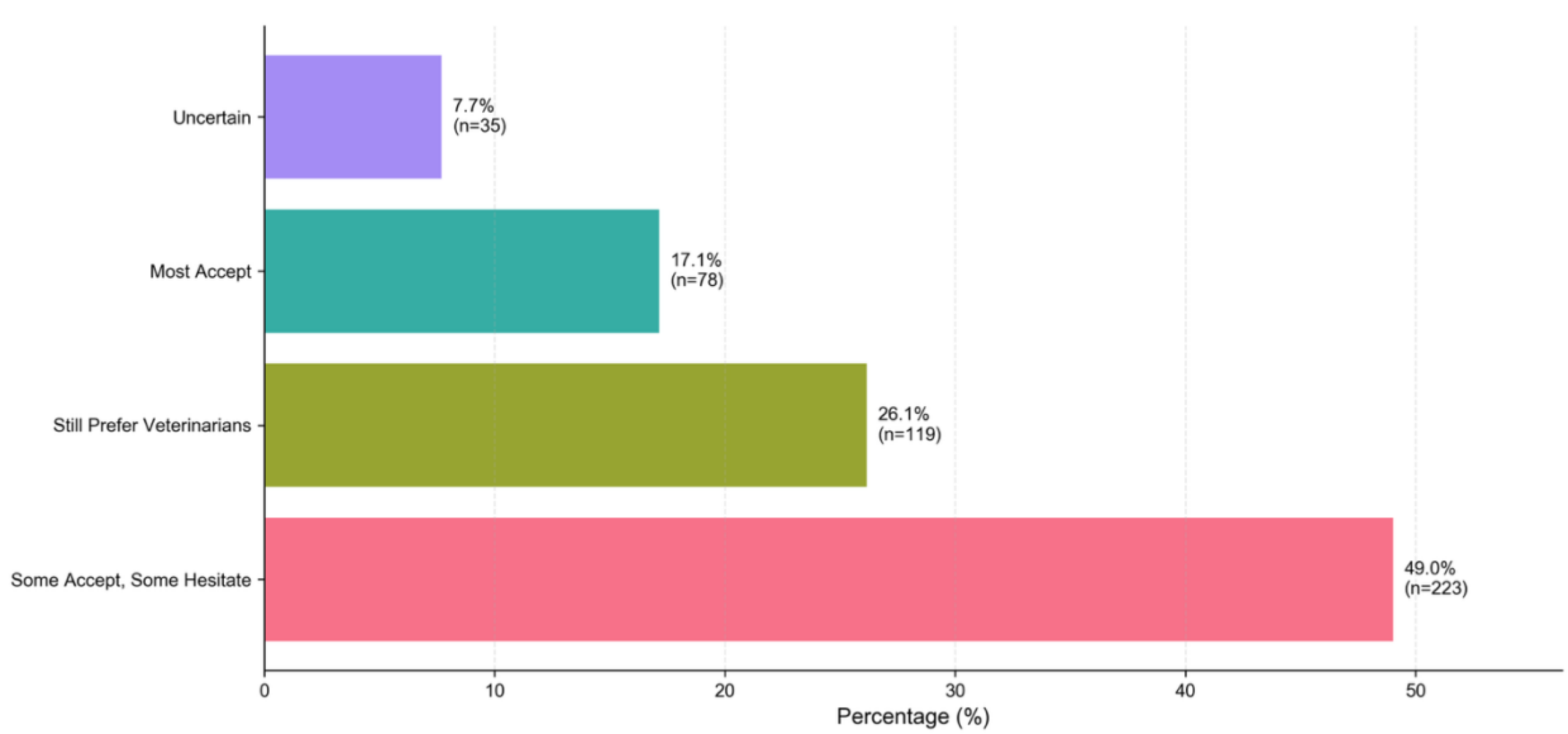
Do You Think Clients Will Accept AI-Assisted Diagnosis/Treatment?
Uncertain
Most Accept
Still Prefer Veterinarians
Some Accept, Some Hesitate
7.7% (n=35)
17.1% (n=78)
26.1% (n=119)
49.0% (n=223)
0
10
20
30
40
50
Percentage (%)

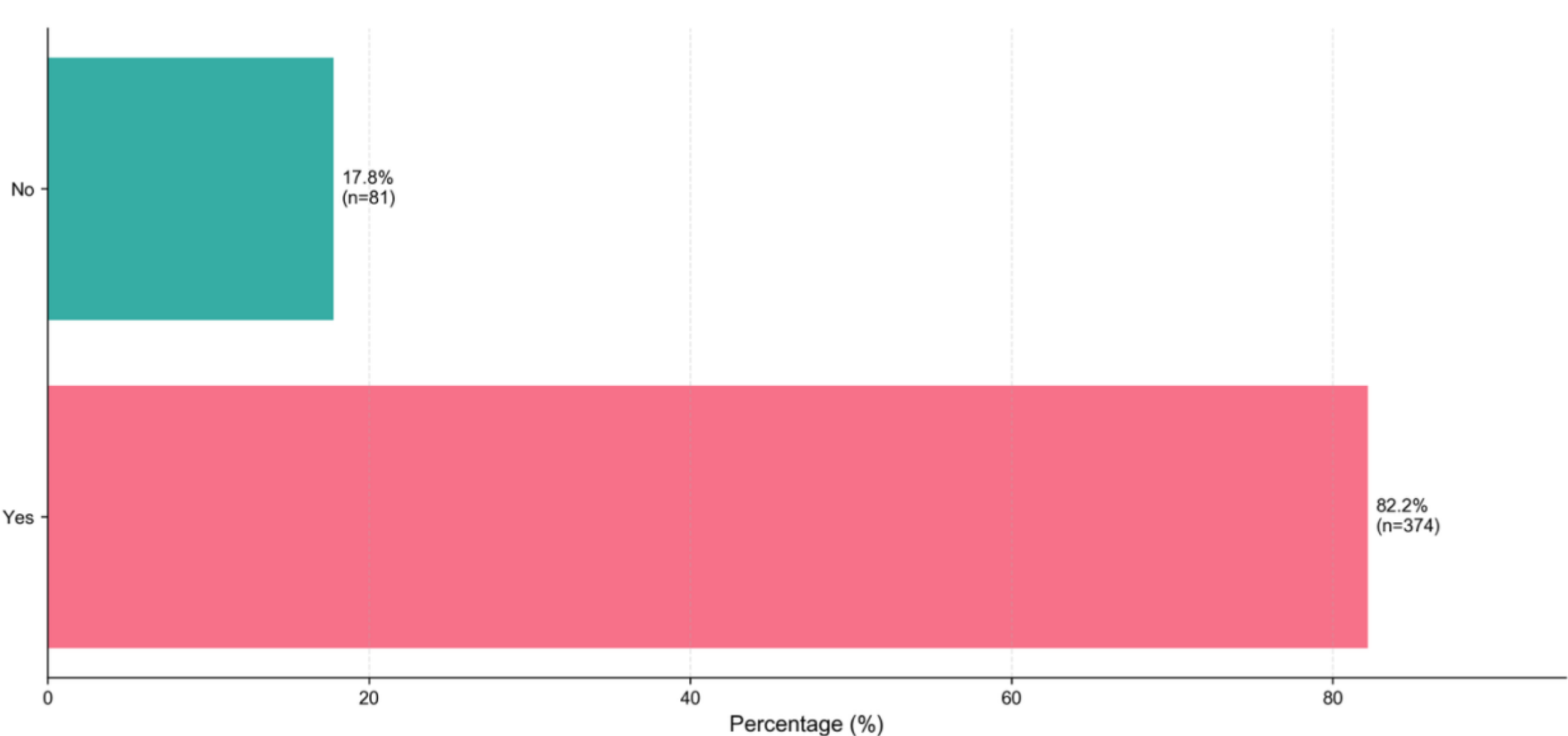
Can AI tools boost a clinic's competitiveness?
No
17.8%
(n=81)
Yes
82.2%
(n=374)
0
20
40
60
80
Percentage (%)

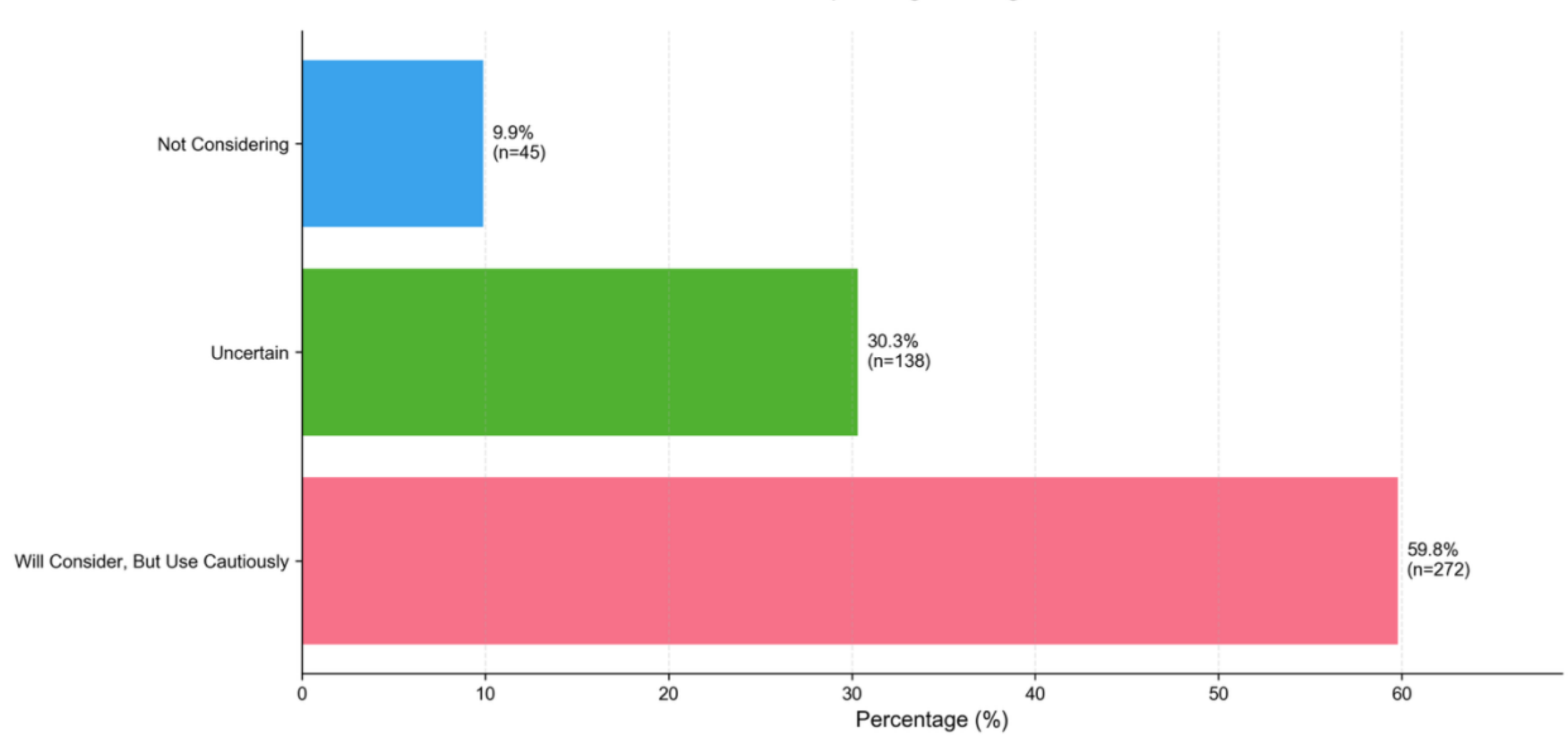
Will You Consider Expanding AI Usage in Your Clinic Soon?
Not Considering
9.9%
(n=45)
Uncertain
30.3%
(n=138)
Will Consider, But Use Cautiously
59.8%
(n=272)
0
10
20
30
40
50
60
Percentage (%)

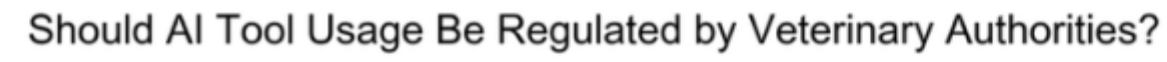
Should AI Tool Usage Be Regulated by Veterinary Authorities?

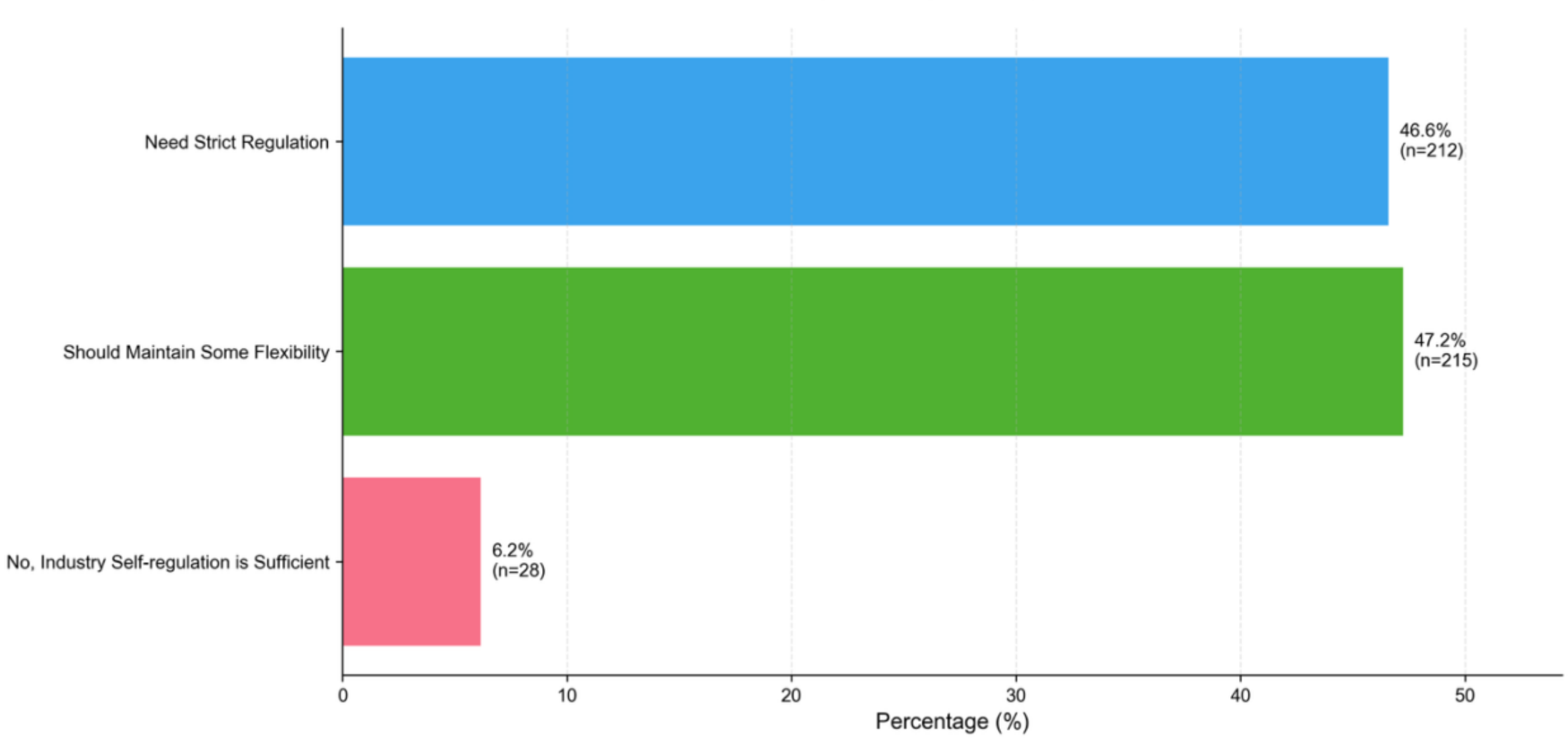